 \documentclass[twocolumn,emulateapj,trackchanges]{aastex61}

\usepackage{graphicx}  

\usepackage{amsmath}
\usepackage{amssymb}
\usepackage{multirow}
\usepackage{color}

\newcommand{\gc}{$\gamma$\,Cas}

\newcommand{\xr}{X-ray}

\def\gtrsim{\mathrel{\hbox{\rlap{\hbox{\lower4pt\hbox{$\sim$}}}\hbox{$>$}}}}
\def\ltsim{\mathrel{\hbox{\rlap{\hbox{\lower4pt\hbox{$\sim$}}}\hbox{$<$}}}}
\definecolor{gray}{rgb}{0.5,0.5,0.5}

\received{January, 4, 2021}   
\submitjournal{ApJ}

\shorttitle{Automated photometry of $\gamma$ Cassiopeiae}
\shortauthors{Smith \& Henry}

\begin{document}

\title{Automated photometry of $\gamma$ Cassiopeiae: the last roundup}

\author{M. A. Smith}
\affiliation{NSF's National Optical-Infrared Astronomy Research Laboratory,
950 N. Cherry Ave., Tucson, AZ 85721, USA }

\correspondingauthor{M. Smith}
\email{myronmeister@gmail.com}

`\author{G. W. Henry}
\affiliation{Center of Excellence in Information Systems, Tennessee State
University, Nashville, TN 37209, USA } \\


\begin{abstract}
$\gamma$\,Cas (B0.5IVe) is the noted prototype of a subgroup of classical Be stars exhibiting hard thermal X-ray emission. This paper reports results from a 23-year optical campaign on this star with an Automated Photometric Telescope (APT). A series of unstable long cycles of length 56--91 days has nearly ceased over the last decade. Also, we revise the frequency of the dominant coherent signal at 0.82238\,d$^{-1}$. This signal's amplitude has nearly disappeared in the last 15 years but has somewhat recovered its former strength. We confirm the presence of secondary nonradial pulsation signals found by other authors at frequencies 1.24, 2.48, and 5.03\,d$^{-1}$. The APT data from intensively monitored nights reveal rapidly variable amplitudes among these frequencies. We show that peculiarities in the 0.82\,d$^{-1}$ waveform exist that can vary even over several days. Although the 0.82\,d$^{-1}$ frequency is near the star's presumed rotational frequency, because of its phase slippage with respect to a dip pattern in its far-UV light curve, it is preferable to consider the UV pattern, not the 0.82\,d$^{-1}$ signal, as associated with rotational modulation. We also find hints of the UV dip pattern in periodograms of seasonal data early in our program. 
\end{abstract}

\keywords{Stars: individual --- Stars: emission line, Be --- X-rays: stars }

\section{Introduction}
\label{intrdn}

As the prototype of the classical Be stars, \gc\ (B0.5\,IVe) 
has exhibited a long history of multiwavelength variability, ranging 
from disappearance of its Be decretion disk to optical and X-ray flaring
\citep[][``S19'']{Doazan et al.1983,Harmanec2002,Smith2019}
Apart from its optical variability, it has
become an object of interest among high-energy
astronomers since its discovery as a hard X-ray 
emitter \citep[][]{Mason et al.1976,White et al.1982}.
Given these discoveries, \gc~has been a target of several
multiwavelength campaigns.

With a considerable X-ray flux (L$_x$/L$_{\rm bol}$ = 3-5$\times$10$^{-6}$),
it is also the prototype of an X-ray class of at least 25 members
\citep[][``N20a'']{Naze et al.2020a}.
These stars are defined by their hard (but thermal) X-ray spectra, which 
exhibit emission lines from  multiple thermal components.
The spectrum of \gc~ itself indicates a dominant $kT$ 
$\approx$ 14\,keV plasma that overwhelms fluxes at all
\xr\,energies. Its X-ray light curve is variable over timescales 
from seconds to more than a year. A peculiar, if not unique,
characteristic of the apparently bright members of the class 
is the existence of ubiquitous rapid X-ray``quasi-flares". 
These features have decay times as short as 4\,s, proving
they are formed in photospheric densities 
\citep[][``SRC'']{1998aApJ..503...877S}. 
A review of the properties of these unique 
X-ray Be stars is given in \citet[][``SLM2016'']{Smith et al.2016}.

To bridge the optical and X-ray domains, SRC and
\citet[][``SRH'']{Smith et al.1998b} were able to
conduct a 34-hour time series with the 
Short-Wavelength Prime (SWP) camera of the International 
Ultraviolet Explorer (IUE) in 1996 January and also
simultaneous 21.5\,hour monitoring with the Hubble Space 
Telescope/Goddard High Resolution Spectrometer (GHRS)
and the Rossi X-ray Timing Explorer (RXTE) on 1996 March 14-15. 
With the decommissionings of the IUE and GHRS, no further UV monitoring has
been possible.

   Several interesting results came of these campaigns,
notably flux correlations between the optical and X-ray domains.
GHRS spectra were binned in wavelength
to construct a high SNR and high time-resolution quasi-continuum 
``UVC" light curve. Two prominent ($\sim$1\%) dips, each
lasting a few hours and separated 
by about 9 hours, were visible. Photospheric UV spectral lines
of Si$^{2+}$, Si$^{3+}$, Ni$^{+}$, Fe$^{+}$, Fe$^{4+}$, and S$^{3+}$
in the same dataset showed correlations/anticorrelations with the UVC
curve \citep[][``SR99''; Smith \& Robinson 2003]{Smith&Robinson1999}.
These UV diagnostics were in turn correlated with the
simultaneous X-ray fluxes. To add to this mix, we found 
the two dips separated by 9\,hours observed by the
IUE 57 days earlier. The same dip pattern emerged from IUE
observations in 1982 found in the IUE archive. 
Although the 
$\lambda\lambda$1407--1417 GHRS light curve could not exhibit color 
changes during the events, the broad wavelength range of IUE/SWP
spectra revealed that the color changes
were consistent with absorptions by large cool clouds 
attached to the star over intermediate latitudes (SRH). 

Another pattern in the GHRS data was
short-lived {migrating subfeatures {\it (msf)} 
that moved blue-to-red across line profiles.
Similar features had been found in optical spectra by
\citet[][]{Yang et al.1988} and \citet{Smith1995}.
Because this phenomenon had been observed only spectroscopically 
for \gc,\,we were curious to see whether it also has a signature 
in broad-band photometry.

These patterns indicated the need for long-term monitoring
of the star. Therefore, in 1997 we initiated a campaign with 
Tennessee State University's T3 Automated Photometric Telescope (APT). 
Although the APT shared observing time with several other scientific projects, 
we planned to observe \gc\,a few times on every available photometric night 
it was in view. Occasionally we could dedicate full nights to the program.

The first discovery from the APT campaign
was of noncoherent, unstable ``long cycles" 
ranging in length from about 56 to 91\,days
\citep[][``RSH''; Smith et al. 2006 (Paper\,1); 
Henry \& Smith 2012 (Paper 2)]{Robinson et al.2002}. 
The amplitudes of the cycles observed in $V$ are 
often larger than in $B$, which implies they are
caused by density modulations within the decretion disk.

To see if these long cycles were related
to the star's X-ray flux, RSH requested and were
granted six 27-hour RXTE visits during 2000--2001.
The exposure durations were chosen to average the flux over 
the star's estimated rotational period.  Intervals between 
successive visits were doubled, such that RXTE 
covered a timescale range from a week to almost 11 months.
Paper\,1 reported that X-ray and optical fluxes during this 
interval revealed a sinusoidal fit to the APT
long-cycle variations that, when suitably scaled to the X-ray fluxes,
showed a very good match. They suggested that the long cycles in the
optical and X-ray regimes were caused by a magnetorotational disk dynamo.
The authors extended this correlation 
by demonstrating a reasonably good  prediction of the X-ray flux from 
the ongoing APT monitoring. 
An updated APT dataset was investigated by \citet[][]
{Motch et al.2015} using the  RXTE All Sky Monitoring
data as well as a later Japanese MAXI (X-ray) dataset. 
These authors confirmed the optical/X-ray correlation of Paper\,1 and 
found that the APT and X-ray datasets correlated from short (a few 
hours) to very long (weeks or longer) timescales as well. 
The latter X-ray datasets were independent of the X-ray data in Paper\,1.
Also, these authors found that
there is no visible time lag between the two sets of variations, in
contrast to lags typical of X-ray Be-NS (neutron star)  binaries. 

In addition to long cycles, Papers 1 and 2 reported a coherent signal 
with $P$ = 1.2158\,d (0.82\,d$^{-1}$$\equiv$``$f_{82}$"). We had noticed 
that this frequency is consistent with the expected rotational period
of $\sim$1.24\,d, according to our estimated physical parameters for
the star. Thus, we adopted 1.21\,d as the
rotational period. Similarly, we suggested that the periodic UVC
dips were caused by rotational modulation of anchored clouds.
This paper will revisit these findings.

\section{A sketch of suggested X-ray mechanisms}

As it is relevant to the APT study,
we sketch a history of attempts to explain the 
production of hard, thermal X-ray flux in \gc\,and its association 
with certain optical and UV variabilities.

\citet[][]{Harmanec et al.2000} discovered that \gc\,is 
a binary system. Its orbital period is 203.5\,d, and it
is in a nearly circular orbit ($e$ $\ltsim$0.03). 
Although the secondary's mass is 0.9$\pm{0.1}$$M_\odot$ 
\citep[][Smith et al. 2012 ``SLM'']{Nemravova et al.2012}, 
its evolutionary status is unknown. However, the evidence 
is strong that the Be star is a blue straggler 
\citep[][2017b]{Mamajek2017a} and therefore probably has an 
envelope-stripped or degenerate secondary.  Assuming the secondary's 
mass estimate is accurate, the mass is too low for it to be a NS, 
but it is appropriate for a white dwarf (WD) and (envelope-stripped) 
helium stars. The limit straddles the masses of sdO stars.

 \citet[][]{Wang et al.2017} have cross-correlated IUE spectra of \gc\,but
found no evidence of a far-UV contribution in the far-UV 
down to a level of 0.6\%. This result rules out a range of types
of evolved, low-luminosity  companions. However, \citet[][]{Wang et al.2021} 
performed the same tests on 13 other early-type Be stars not previously
known to be in binaries and found 10 from this sample exhibit at least traces 
of a spectrum of a hot subluminous secondary such as an sdO star.
Using the same technique, 
\citet[][]{Gies et al.1998} and  \citet[][2013, 2016]{Peters et al.2008}
had previously discovered substantial UV contributions from 
sdO secondaries of three Be binaries ($\phi$\,Per,
FY\,CMa, and 59\,Cyg). In a fourth case, HR\,2142 only a faint contribution 
can be seen from a ``sdO in transition." Clearly, sdO's are the most 
likely kind of secondary in early-Be systems. Thus,
to see if these sdO's could be seen against \gc\,as 
the primary star, we substituted the appropriate physical parameters of 
\gc\,for the parameters of the actual primaries and recomputed the secondary 
flux contributions to simulated \gc-sdO systems. This exercise confirmed 
that sdO contributions would still be recognizable for the first three 
imaginary cases if \gc\,had been the primary.
We add finally that at least two of the binaries in the Peters-Gies et al.
sample, 59\,Cyg and $\phi$\,Per (those with the brightest and most massive 
hot secondaries) are faint and soft X-ray systems
\citep[][ HEASARC\, Rosat\,Data\,Archive2020]{Naze et al.2020a}, which
clearly cannot be confused with \gc~star emissions.

Even before the binarity of \gc\,was discovered, 
accretion of Be wind onto a degenerate object 
was suggested as the source of the hard X-ray emission. 
\citet[][]{White et al.1982} and \citet[][]{Murakami et al.1986} had argued 
that a secondary ought to be a NS or WD.
More recently, various authors have again suggested 
accretion involving degenerate or hot secondaries:  
a WD \citep[][]{Hamaguchi et al.2016}, a NS in propeller stage
\citep[][]{Postnov et al.2017}, or interactions
between an sdO wind and the Be disk \citep[][``L20'']{Langer et al.2020}. 


Difficulties described by SLM16
in reconciling the unique \xr~characteristics of \gc\,with 
those of other X-ray classes of Be stars 
motivated SRC, SR99, and \citet[][``RS00'']{Robinson&Smith2000} to advance 
a very different mechanism for the hard-\xr\,production:
the star-disk magnetic interaction hypothesis. 

This idea requires the tangling of field lines from 
putative small-scale magnetic surface complexes 
\citep[e.g.,][]{Cantiello&Braithwaite2011} and a 
toroidal field in the inner decretion disk. The different angular
rotation rates of the star and the disk cause interactions of star-disk 
fields, which in a short time entangle,
break, reconnect, and ultimately relax. This process releases
magnetic energy, which accelerates embedded particles
in high-energy beams. Some of these are guided by 
protruding surface field lines toward the star. In fact, the existence of
downstreaming matter can be inferred from highly redshifted 
absorption lines in the 1996 March GHRS time series (SR99). 
According to simulations by RS00, nearly 
monoenergetic (200\,keV) electron beams impact and thermalize at 
the surface, causing local explosions (also called  ``flares"). 
Their detritus accumulates in low-density canopies and decays in 
$\sim$20 mins,  producing a basal flux of the same high temperature.

\section{Observations}
\label{obbs}

We have been conducting Johnson $B$ and $V$ photometric observations of 
\gc~since 1997 with the T3 0.4~m APT facility
at Fairborn Observatory in southern 
Arizona \citep[][b, Henry 1999, Eaton et al. 2003]{Henry1995a}.  The 
APT acquires successive brightness measurements of individual target 
stars with a single-channel photometer using a 
temperature-stabilized EMI 9924B photomultiplier tube.  Each 
observation of \gc,\,which we refer to as a {\em group} observation, consists 
of the mean of three $B$ and $V$ differential measurements of $\gamma$~Cas with 
respect to its comparison star (HD\,6210) and the mean of two measurements
of the check star (HD\,5395) with respect to the 
comparison star. 

To avoid saturating the detector on 2nd magnitude \gc,~ 
we used a 3.8~mag neutral density filter to observe all three stars, with the
exception of the 1997 season when we used different neutral 
density filters for \gc\,and the comparison stars.

\gc\,comes to opposition in early October, and our APT observing seasons
always covered the months of September through early February. Telescope
closure was generally forced by Arizona's summer monsoons.  However, if 
the monsoons arrived late, a new season included some nights 
in June-July.  In the analysis that follows, we label each season 
by the year corresponding to the star's opposition.  Our final 
dataset covers 23 consecutive observing seasons from 1997 through 2019. 

On most clear nights the APT was programmed to acquire 1--4
observations spaced by $\ge$2 hours.  During most observing seasons, 
a few nights near opposition were dedicated to monitoring for several 
hours. On those nights the APT acquired 1-4 group observations 
spaced about 8 minutes apart.  On good nights the external precision 
of the group means was typically 0.003--0.004~mag, as determined from
observations of pairs of constant stars.  Since the APT can acquire 
observations in marginal photometric conditions, we rejected as outliers 
any group mean differential magnitudes with standard deviations greater
than 0.01\,mag.  Because we used different neutral density filters for the 
variable and comparison stars in the 1997 season, we were forced to adjust 
those means to the 1998 season means.  Finally,
the check minus comparison star differential magnitudes demonstrated that 
both are constant to $\le$0.005 mag on seasonal and year-to-year
time scales.

In this paper we present our final 1997--2019 dataset for \gc,\,consisting 
of 5554 observations in the $B$ and 5488 in the $V$ passbands.
We have cleaned the dataset by fitting sine curves to the long cycle 
($\sim$70-80 day) variability in single observing seasons and rejected as 
outliers those observations with residuals from the least-squares sine fit of 
$\ge$2.5$\sigma$.  For several of the later observing seasons, where
we could not obtain a reliable long cycle, we rejected observations that
were $\ge$ 3.0$\sigma$ from the seasonal mean magnitudes.

Our analysis technique employed the frequency-search method of
\citet[][]{Vanicek1971}, based on least-squares fitting of sine curves, to 
search for periodicities in various combinations of the yearly photometric 
datasets. This method uses the reduction factor in the data's variance 
as a goodness-of-fit parameter. For all analyses we initially scanned trial 
frequencies over a range 0.005--6.0\,d$^{-1}$.  Formal uncertainties in the 
best-fit periods and amplitudes were computed from standard propagation
formulae.  We consider a signal to be real only if it is found in both $B$
and $V$ datasets.  For multiseasonal analyses we forced the seasonal means
to the same value to eliminate those low-frequency variations that can be 
seen in the long-term light curve (Fig.\,1). If $>$3$\sigma$ outliers were
present at any of the frequencies detected within the individual observing
seasons, we removed those observations from our dataset and recomputed all 
frequencies for that season.  In total, we rejected $\approx$6\% of the 
observations acquired by the APT.  The total number of observations given 
above are the observations that survived analyses of the individual observing 
seasons. These data are listed in Table\,1.  The full table is given on 
the VizieR website accompanying this paper.
The mean $V$ and $(B-V)$ magnitudes in the Johnson system 
are $<V>$ = 2.165 and $<B - V>$ = -0.083.

\begin{table}
\begin{center}
\caption{Automatic Photoelectric Telescope Observations of \gc~ (Seasons 1997-2019)}
\begin{tabular}{@{}ccrrl@{}}
\hline
\hline
Date & Var $B$ & Var $V$ &  Chk $B$ & Chk $V$ \\
(RJD) & (mag) & (mag) & (mag) & (mag) \\ 
\hline
50718.6953 & -4.306 & -3.671 & -0.809 & -1.213 \\
50718.9258 & -4.306 & -3.672 & -0.805 & -1.219 \\
50720.7930 & -4.299 & -3.674 & -0.812 & -1.215 \\
50720.9180 & -4.300 & -3.668 & 99.999 & -1.220 \\
\hline
\end{tabular}
\end{center}
Note: (Stub.)

The full dataset may be retrieved from the Journal's VizieR website
for this paper.
A value of 99.999 indicates the differential magnitude
had an uncertainty $>$0.01~mag and was rejected.
\end{table}

\section{Results}

\subsection{The full light curve }
The $B$ differential magnitude dataset is plotted in 
Fig.\,1.  Besides short-term brightness variations, the plot 
exhibits a sinuous character over a range of 
$\sim$0.05\,mag in the 23 seasonal means. The bottom curve shows
that the $B-V$ color index is reddened by 0.04 mag during this
time, though not in strict correlation with $B$ (or $V$) magnitudes. 
Because there is no other plausible source of surplus red continuum flux in 
the \gc\,system, we may assume that the variable reddening results from 
changes in disk extent and/or density.
However, we also notice the unusual variations occurring between Seasons 2018 
and 2019 when the $B$ flux has decreased and the $V$ flux decreased too, 
but less so. It is possible that the disk has developed enough
in this time that it has become opaque even in the blue.
Then, assuming the inner disk edge 
occults part of the star, its enhanced optical thickness will 
dim the combined star/disk light even in the $B$ passband.

\subsection{Long cycles}

Table\,2 lists by APT season the numbers of $B$ and $V$ observations and the
lengths of the long cycles. These are taken from Papers\,1 and 2 and, from 
Season 2012 on, new observations. 
As previously noted, the waveforms for the long cycles are not
always sinusoidal. They can grow, damp out, or exhibit a net trend. 
At times they ``morphed" to a new quasi-period within two weeks or less. 
Errors in the long-cycle lengths were estimated in Papers 1 and 2 to be
${\pm 1}$ day for simple sinusoids to ${\pm 2}$ days for damped cases. 
In contrast to Paper\,2, the full (peak-to-peak) 
amplitudes we list in Table\,3 were computed by sinusoidal fits and without
(sparce) summer observations.
Thus the amplitudes here are not identical to those in the previous papers.
The errors in cycle amplitudes are likewise dependent on the character of 
the variations and therefore also difficult to assign. We estimate them 
conservatively to be ${\pm 15}$\% (see Fig.\,2). The full-amplitude 
detection threshold is about $\sim$1\,mmag. 

\begin{figure} 
\begin{center}
\includegraphics*[width=6.2cm,angle=90]{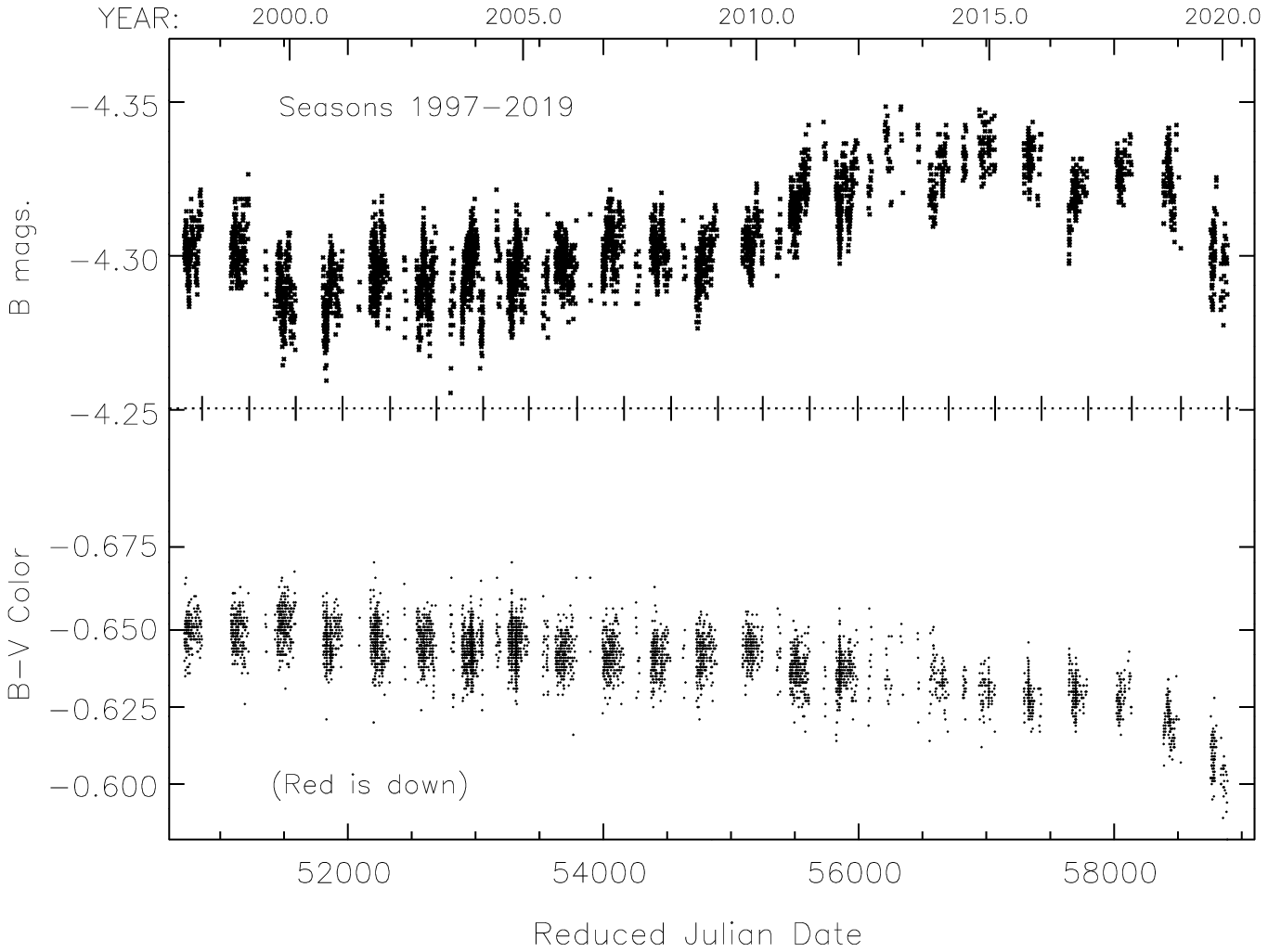}
\end{center}
\caption{The APT $B$ and $B-V$ magnitudes for the \gc\,program
for Seasons 1997--2019. Tick lines mark each season's end. 
The brightening and reddening segment after RJD\,55400 marks an 
outburst during 2010-2011. In the last season the $B-V$ color reddens
because the Be star fades more in $B$ than in $V$. }
\end{figure}

\begin{table}
\begin{center}
\caption{Summary of APT Observations with long period and $f_{82}$ properties}
\begin{tabular}{lr|cr|r}
\hline 
\hline
 Season & \# Obsns. &  Long Cycle &  Long Cycle &  $f_{82}$~~~~~~~~  \\
     & $B$/$V$~~~  & ``Periods"     & B/V Ampls. &  $B/V$\,Ampls. \\
\hline
1997 & 179/183  & 61 & 14.0/15.9 &  4.4/4.4  \\  
1998 & 206/209  & 65 & 6.8/7.6   & 5.9/6.1  \\  
1999 & 254/251  & 72 & 14.1/14.9  & 4.8/6.1   \\ 
2000 & 290/290  & 91 & 14.8/17.6  & 3.4/6.1   \\ 
2001 & 332/327  & 73 & 10.6/11.9  & 7.1/6.4 \\ 
2002 & 300/300  & 80 & 16.4/21.0  & 7.6/4.7   \\ 
2003 & 659/655  & 90: & 19:/21:$^3$ & 6.3/7.2   \\ 
2004 & 647/641  & 85 & 11.2/15.8  & 3.6/2.7 \\ 
2005 & 287/275  & 66 & 6.0/4.9    & 5.3/2.6   \\ 
2006 & 266/270  & 88 & 13.3/17.6  & 0.6/2.3   \\ 
2007 & 254/248  & 88 & 11.2/13.9  & 0.8/2.0   \\ 
2008 & 245/242  & 60 & 10.8/9.3   & 2.3/2.0   \\ 
2009 & 192/188  & 70 & 10.2/9.3   & 0.7/1.5   \\ 
2010 & 278/278  & 72 & 13.2/19.1  & 1.8/2.8   \\ 
2011 & 326/318  & 73 & 17.7/20.1  & 1.5/2.0   \\ 
2012 &  40/34  & 70 & 10.1/18.3  & 0.0/0.0   \\ 
2013 & 93/93    & -- &  --        & 1.4/0.0   \\ 
2014 & 91/89    & -- &  --        & 3.3/3.4   \\ 
2015 & 132/126  & -- & --         & 3.4/2.9   \\ 
2016 & 173/169  & -- & --         & 2.5/3.0   \\ 
2017 & 87/88    & -- & --         & 2.2/0.0   \\ 
2018 & 120/113  & 56 & 1.8/1.0    & 2.1/1.6   \\ 
2019 & 103/101  & 73 & 1.6/1.9    & 1.5/2.4   \\ 
\hline
\end{tabular}
\end{center}
Notes: (1) For conciseness $B$ and $V$ properties are separated
by a slash symbol. \\
(2) Full amplitudes are in mmag; cycle lengths in days. \\
(3) Paper 1 showed damping/regrowth of cycle amplitude. 
\end{table}

\begin{figure} 
\begin{center}
\vspace*{-0.15in}
\includegraphics*[width=6.2cm,angle=90]{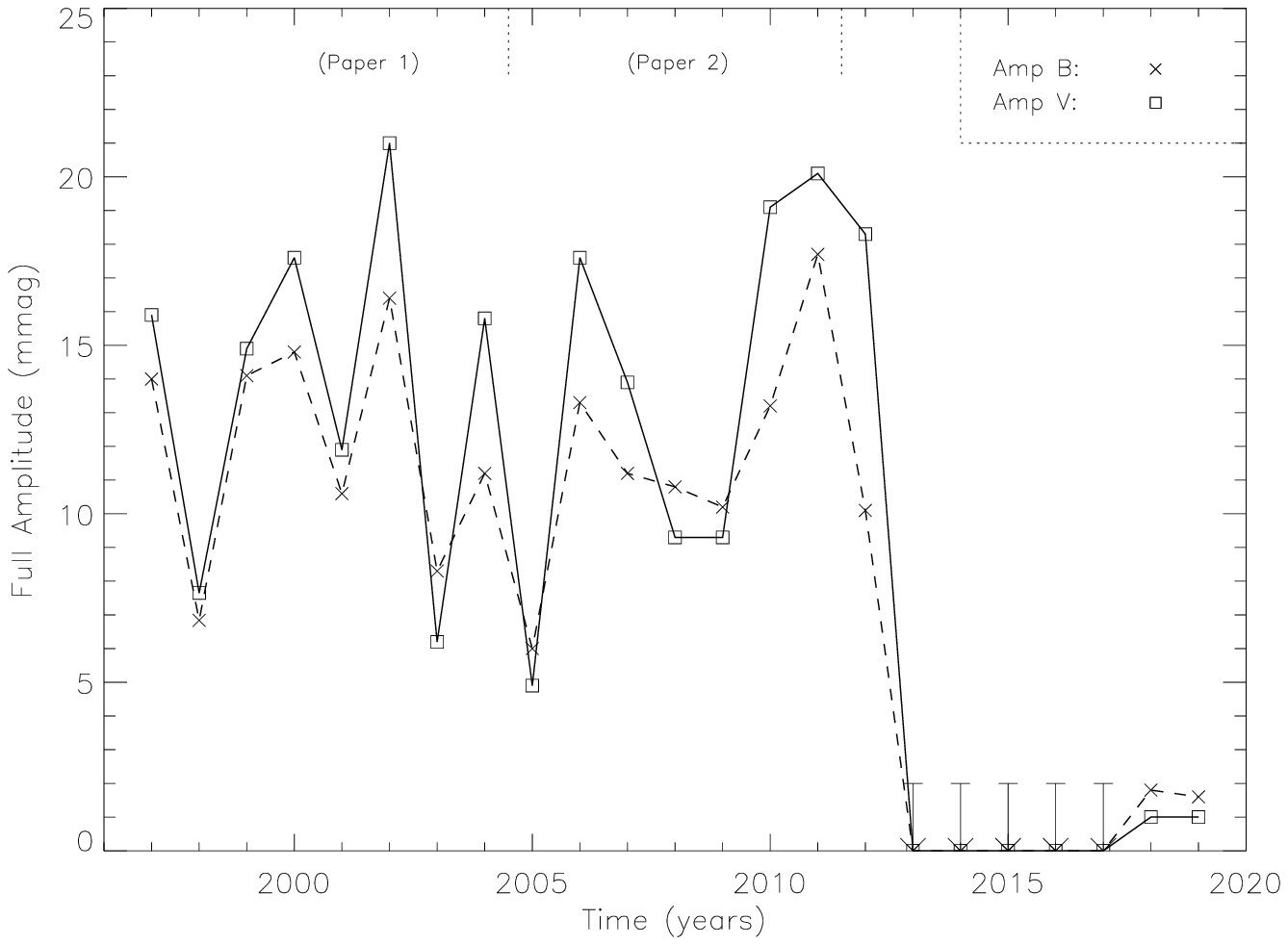}
\end{center}
\caption{ The seasonal APT history of long-cycle
full-amplitudes of \gc.\,Solid and dashed lines and symbols annotate filter. 
The seasons covered by Papers\,1 and 2 are indicated. 
For most of the last several years the amplitudes  have
been too small to detect.  }
\end{figure}

\begin{figure} 
\begin{center}
\vspace*{-0.15in}
\includegraphics*[width=6.2cm,angle=90]{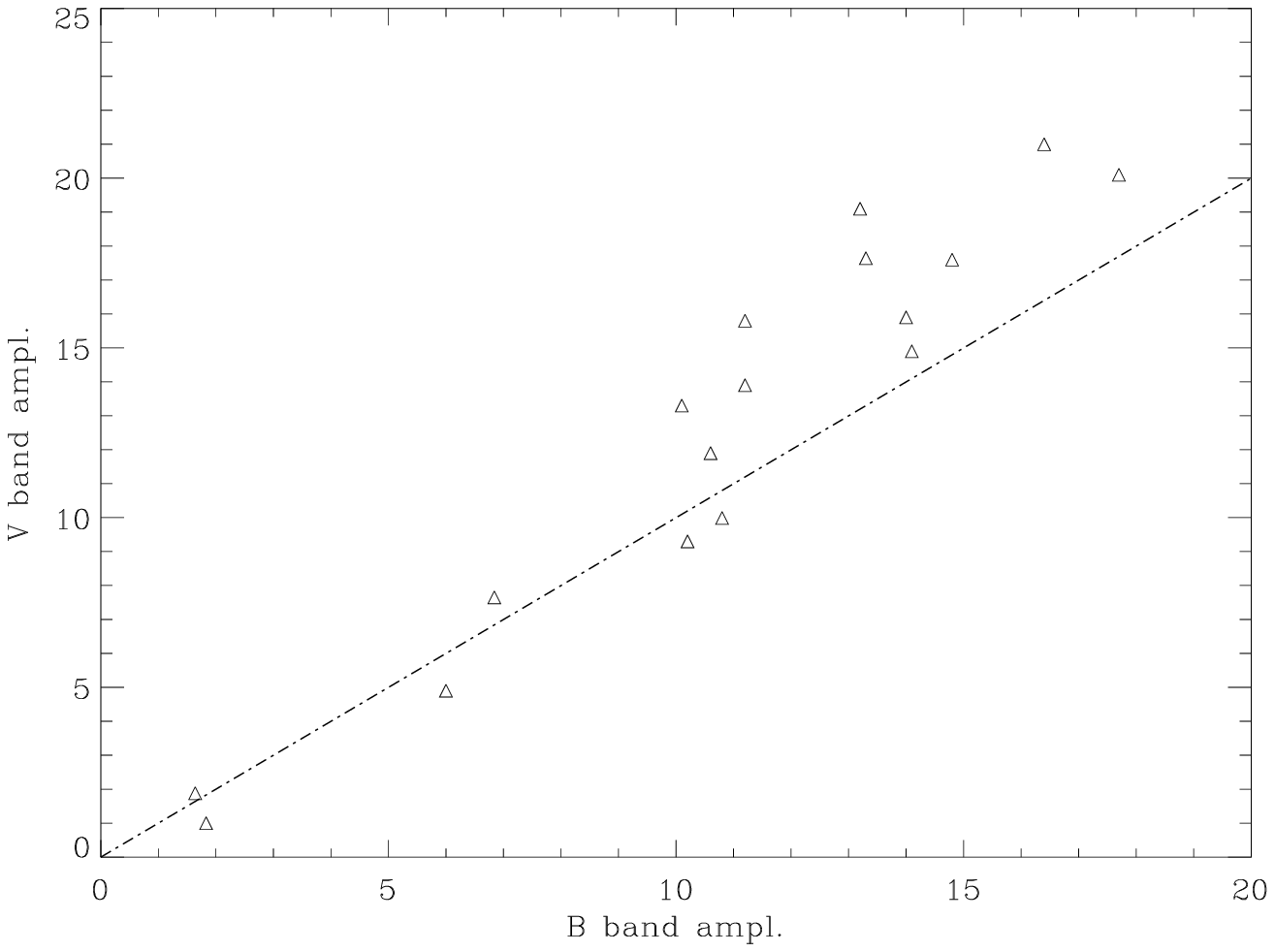}
\end{center}
\caption{The correlation of $B$ and $V$ amplitudes in mmag from the long-cycle
history in Fig.\,2. For amplitudes greater than 10 mmag the $V$ amplitudes
become significantly larger than in $B$, demonstrating that the origin
of these variations is the Be disk.
}
\end{figure}

From the seasonal history of the long cycles given in Fig.\,2 
and the table, their character appears chaotic, offering no obvious 
predictive power or memory of previous cycles. 
Following the cycles' decline in 2012-2013, one sees 
that they may have recovered slightly in Seasons 2018--2019.
This is at least consistent with the long-period, meandering character of
the TESS satellite\footnote{The Transiting Exoplanet 
Survey Satellite (TESS) \citep[][]{Ricker et al.2015} 
was launched by NASA in 2018
to survey the sky with broad-band optical photometry. 
The time cadence for \gc\,observations was 30\,mins.}
observations during this season, which 
further supports this general picture.

\begin{table*}
\begin{center}
\caption{Signal frequencies (d$^{-1}$) and full amplitudes (mmag)}
\begin{tabular}{@{}l|cccc@{}}
\hline 
\hline
  Multiseason &  0.82  &  1.24 &   2.48 &   5.03  \\
Paper 1: &    &    &   &  \\
1997-2004 (B)    & 0.82245  & 1.24310  & 2.47944 &  5.02903  \\
~~~Full ampl.:          &    5.92${\pm 0.35}$   &  3.00${\pm 0.37}$ &  1.53${\pm 0.38}$    &   3.00${\pm 0.38}$ \\
1997-2004 (V)    &   0.82244 &   1.24304 &  2.47937 &  5.02328 \\
~~~Full ampl.:          &   6.80${\pm 0.39}$     &   3.30${\pm 0.42}$    &   1.66${\pm 0.42}$    &   3.31${\pm 0.42}$    \\
Paper 2 &    &    &   &  \\
2005-2011(B)   &   0.82929 &   1.24171 &  2.47975 &  5.03873  \\
          &   2.28${\pm 0.45}$     &     2.11${\pm 0.45}$  &   2.23${\pm 0.45}$   &    1.54${\pm 0.46}$ \\
2005-2011(V)   &   0.81929  &  1.26146  & 2.47980  &  5.02477 \\
          &    2.47${\pm 0.45}$    &   2.36${\pm 0.49}$     &  1.70${\pm 0.49}$     &   1.34${\pm 0.50}$ \\
New data: &    &    &   &  \\
2012-2019 (B)  & 0.82394 &   1.24263 &  2.47946 &  5.03244 \\
          &  2.32${\pm 0.65}$    &   1.58:/2.37${\pm 0.65}$ &    3.49${\pm 0.64}$   &     2.25${\pm 0.66}$  \\
2012-2019 (V)  &  0.82762 &  1.25951 &  2.46901 &  5.02552 \\
          & 2.46${\pm 0.67}$     & 2.00${\pm 0.68}$       & 2.44${\pm 0.65}$    &   2.54${\pm 0.66}$ \\
\hline
23 seasons$^{(3)}$ &  0.82238(10) &  1.2448(19) &  2.481(11) &  5.027(20) \\
     (B) &   3.51${\pm 0.29}$   &     1.89${\pm 0.29}$ &  
2.42${\pm 0.39}$  &   3.00${\pm 0.38}$ \\   
     (V) & 3.58${\pm 0.28}$    & 2.20${\pm 0.29}$  &    2.15${\pm 0.50}$  &      2.60${\pm 0.41}$ \\
\hline
\hline
$f_{82}$ ephemeris &    &    &   &  \\
from 23 seasons:  & $T_{\circ}:$ (RJD) & P\,(d): & $f_{82}$  &  \\
B  &     51086.602 &  1.215975   & 0.822385  &   \\
V   &    51086.612 &  1.215987   & 0.822377  &   \\
Avg. $T_{\circ}$, ~$P$, ~$f$ &    51086.607(5) &  1.21598(1)  & 
 0.82238(1)   &   \\
\hline
\hline
\end{tabular}
\end{center}
\noindent  Notes: 
(1) The 1.21\,d ($f_{82}$) ephemeris is:
 \(\displaystyle \phi = (T - T_\circ)/2 \pi P \) ; \\
(2) $\phi$ = 0.0 refers to the ``faint star" phase; \\ 
(3) The last digit in both the amplitude and frequency is not significant.
\end{table*}

\subsection{Confirmation of $f_{82}$ and other signals}
\label{f82r}

\subsubsection{Search procedure}

 Our search procedure for coherent frequencies was first 
to run our Van\'{\i}\v cek periodogram generator through a broad
frequency range for each season (and each filter) 
and to tabulate the formal errors in the amplitudes of all significant peaks. 
This procedure worked well for $f_{82}$
and for all but one of the signals we may have found
(see reference to 0.76 d$^{-1}$ in $\S$\ref{suddn}).

 We digress to point out that in their analysis of the TESS light curve
of \gc\ during Sectors 17, 18, and 24, \citet[][``LB21'']{Labadie-Bartz et 
al.2021} have discovered a low-amplitude group of NRP modes,
which they designate as ``group g$_1$.'' However, their amplitudes are
too low to be detectable by the APT, and they flutter 
on an unknown timescale. The more stable, and larger-amplitude f$_{82}$, 
first noted in the APT light curves of Papers\,1 and 2, occurs at the low
frequency edge of the g$_1$ group. Because these modes are so weak and 
well separated from aliases associated with the APT observing windows, 
we believe they are not directly related to f$_{82}$.

 In Paper\,2 we found that 
uncertainties in the full amplitudes for single seasons range
from ${\pm 1.0}$ to $\pm{1.5}$ mmag.  After analyzing the
individual seasons, we ran searches on three groups of seasonal datasets: 
early, middle, and late (see Table\,3), according to the seasons added
to Papers 1, 2, and this work. Initial amplitude errors were formal ones, 
as propagated from the Van\'{\i}\v cek analysis.
The errors for multiseasonal periodograms are lower
than the single-season errors.  The 23-season errors were calculated in
the same matter.

To assess the effect of hypothetically fewer observations than were made, we 
split our
database in two, each comprised of even or odd-numbered points, and reran 
our 23-season analysis on four discovered frequencies discussed below.
As might be expected, the solutions for frequencies and amplitudes typically 
bracketed those computed from the full dataset. The computed r.m.s. values
based on the even minus odd observation differences varied between 1 and  
1$\frac{1}{2}$ times the r.m.s. of the full-set solutions. 
In the lower (``23\,seasons") panel of Table\,3 we have replaced the 
errors of the  Van\'{\i}\v cek solutions with those
from the even/odd comparisons in cases where they were larger.
We also note that the amplitudes computed in the even/odd analyses fluctuated
apparently randomly; there was no trend to smaller or larger values.

We repeated these trials four times by computing results for every {\em fourth}
observation for these frequencies. About half these trials settled on
the correct frequency, within the error windows given in Table\,3. In the
other cases a false peak (usually one of the adjacent annual aliases) 
was chosen, since the correct peak often did not stand out above them. 
With such a high failure rate as this, it was clear that we had reached the 
limit of our ability to detect astrophysical signals.

\subsubsection{Discovery of coherent signals}

The time history of $f_{82}$ full-amplitudes found in APT data is given in 
Fig.\,4 (values for 1997-2011 are from Paper\,2). Although this signal 
was strong in early seasons, the amplitudes have decreased significantly
after 2004-2005 to being barely visible until 2013 or 2014.  From concurrent 
SMEI observations, 
\citet[][``B20'']{Borre et al.2020}
reported a similar decrease in the f$_{82}$ amplitude. 
It appears to have partially recovered in the following few years, 
but it is not visible in the later periodogram obtained during TESS Sector 
17-18 (2019-2020) observations   
\citep[][``N20c'']{Naze et al.2020c}. According to Fig.\,4,
we cannot attest to the nonzero values for Seasons 2017-2019.

The lower panel of Table\,3  gives our $f_{82}$ ephemeris for 
all the data in both filters. 
The agreements between the $f_{82}$ 23-seasons frequency for 
Seasons 1997--2019 and the three multiseason segments suggest that this 
frequency has been coherent from when the APT monitoring began in Season 1997 
through Season 2011 and probably into some late seasons. Our revised frequency
reduces the discrepancy between the values reported in Paper\,2 and B20 
(0.82247\,d$^{-1}$ vs. 0.82215\,d$^{-1}$) 
by ${\frac 13}$. These values now differ by
close to 1.0 cycle over their timespan. It is likely that either they 
or we have miscounted by one cycle over the span of several thousand.

Other than 0.82\,d$^{-1}$, we found multiseasonal
signals near frequencies 1.24\,d$^{-1}$, 2.48\,d$^{-1}$, and 5.03\,d$^{-1}$, 
very similar, though not always identical to, results by B20, N20c, and LB21.
Periodograms for frequencies surrounding these values are exhibited in Fig.\,5, 
and relevant parameters for them are listed in Table\,3. The 2.48\,d$^{-1}$ 
signal seems to be a robust frequency for most, if not all, of the APT 
observing seasons.  The 5.03\,d$^{-1}$ and 1.24\,d$^{-1}$ signals require 
additional notes.

N20c determined a peak value of 5.054\,d$^{-1}$ for f$_{5.03}$, 
which lies at our frequency error limit. Our periodogram 
for this signal shows evidence of stronger annual and daily aliases than
the others. In marginal detection cases like this, these may be due to
the frequency's proximity to the diurnal harmonic at 5.0\,d$^{-1}$.

The f$_{1.24}$ signal (B20's f$_{1.25}$) was the most problematic to 
characterize.  The filter-to-filter disparity in its amplitudes is large 
compared to results for the other signals, making our late seasons' solutions 
for it less reliable. Indeed, this signal weakened in seasons after 2004.
Moreover, we suspect its amplitude was generally highly variable. 
For example, N20c state that in the TESS Sectors 17-18 
they found ``no trace" of B20's f$_{1.25}$. Yet, f$_{1.24}$ and
f$_{2.48}$ signals were both present but variable at least for a few 
days during the beginning of Sector 18, as seen in their Fig.\,6 light curve.
Finally, 
for this event or other times, we did not find that the APT or TESS 
amplitudes of the two frequencies change together,
as would be expected if the two were related through a harmonic resonance. 

\begin{figure} 
\begin{center}
\includegraphics*[width=6.2cm,angle=90]{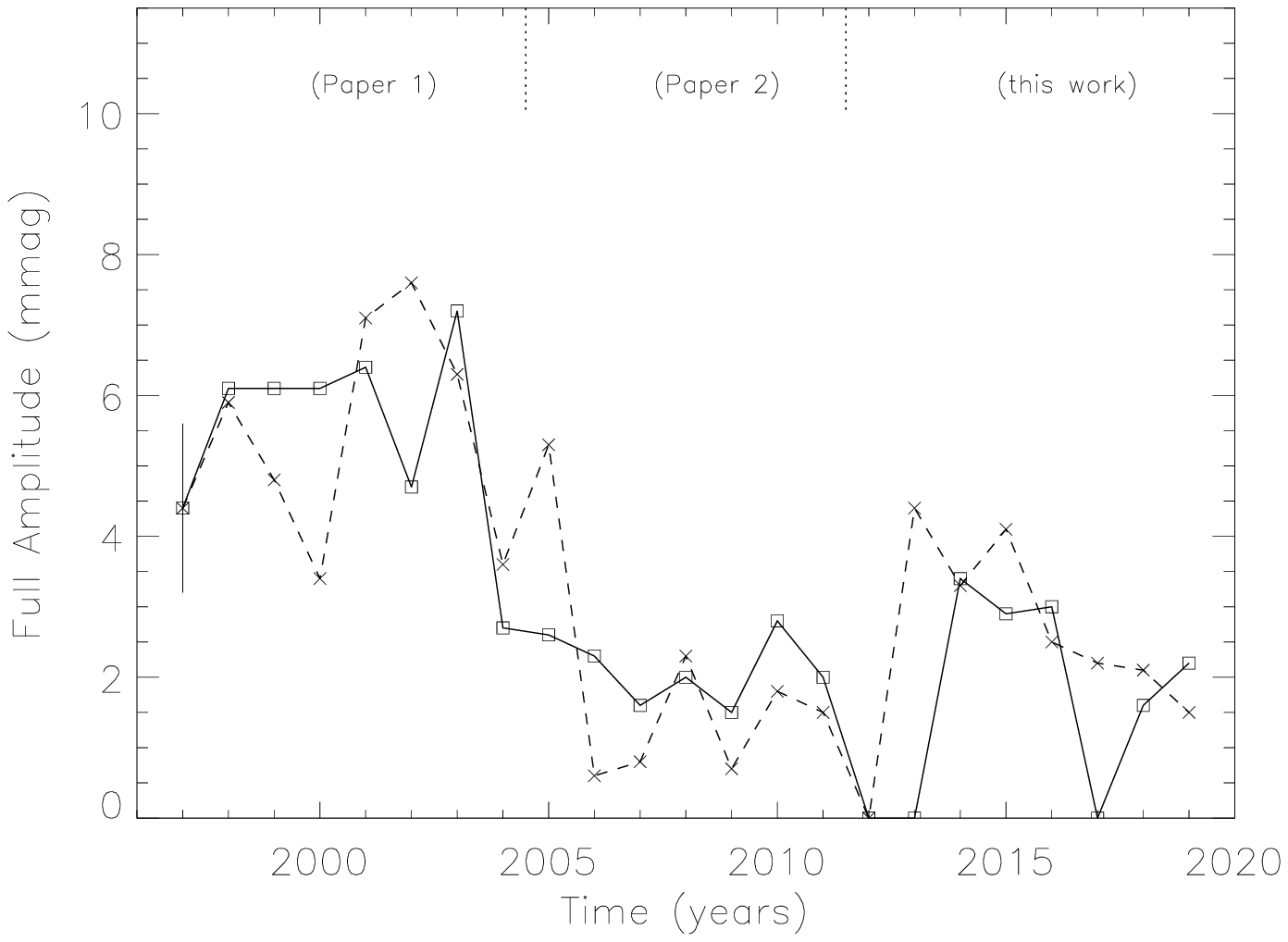}
\end{center}
\caption{The time history of the seasonal full amplitudes for the $f_{82}$
signal for $B$ (X symbols) and $V$ (squares) datasets.  
Errors for the amplitudes of ${\pm 1.2}$ mmag are averages of seasonal 
values given in Paper\,2.}
\end{figure} 

\begin{figure} 
\begin{center}
\includegraphics*[width=6.2cm,angle=90]{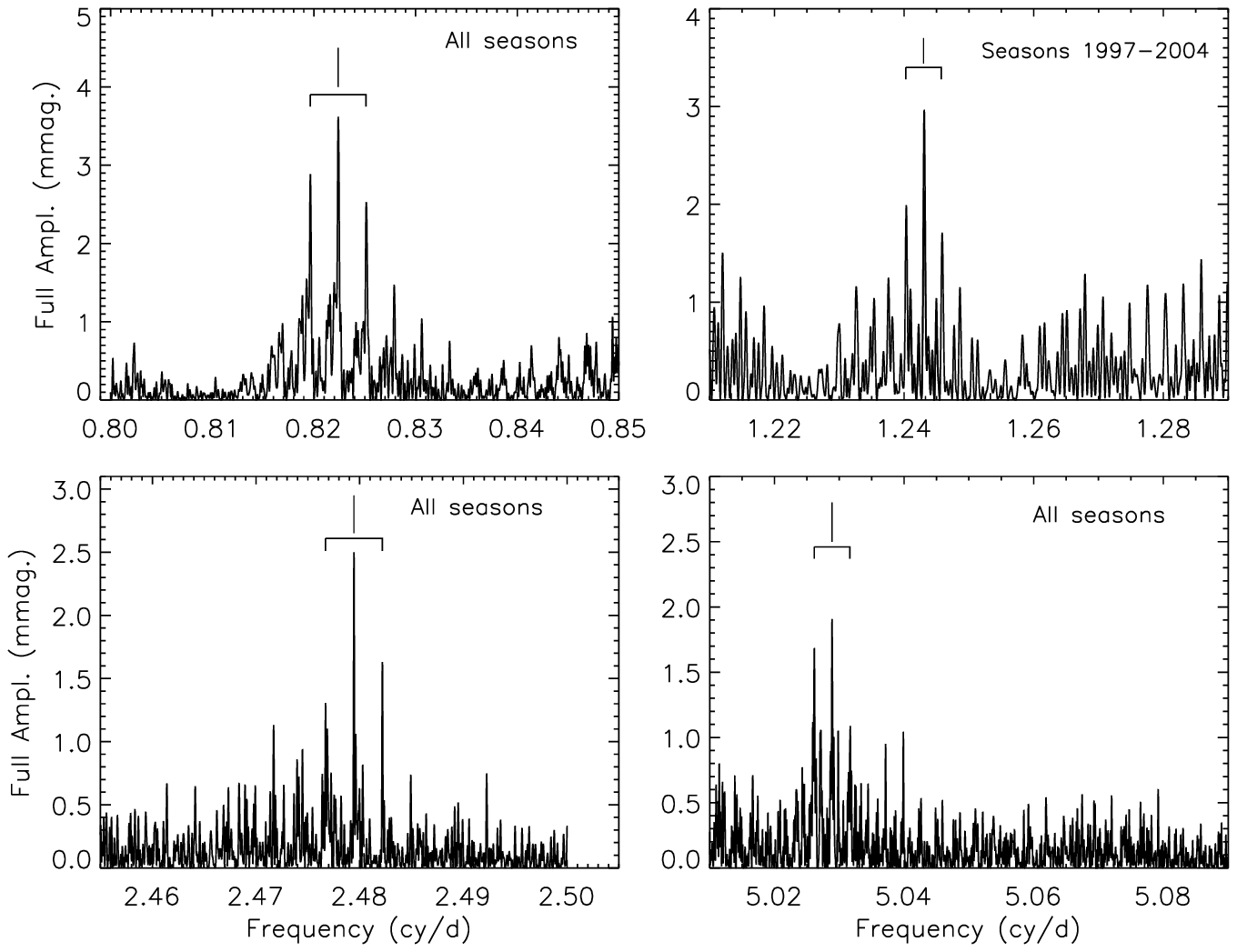}
\end{center}
\caption{The $B$ filter full-program periodograms for the four frequencies
found for most of the APT seasons covered in this paper. The comb astride 
the vertical line denoting these frequencies highlights the ${\pm 1}$ 
annual aliasing pattern.}
\end{figure} 

We searched our periodograms for additional signals (for example, 
the weak signal at 7.57\,d$^{-1}$, found by N20c in a TESS light curve)
but detected none.
In addition to the semi-stable frequencies just noted, we will cite a
short-lived frequency in $\S$\ref{suddn} and discuss an intermittent one
in $\S$\ref{uvcuvc}.

Before proceeding further, notice that there is no
correlation between the amplitudes of the long cycles and the $f_{82}$ signal. 
We note also that according to Table\,1 and \citet{Pollmann2021},
the disk of \gc~has been building since 2000 through early-2021. 
In the middle of this interval, 
\gc~ underwent a Be outburst in 2010--2011 showing increased optical
continuum and H$\alpha$ line brightening. As reported in Paper\,2, during
the initial few weeks of this outburst, an accelerated brightening of the
optical continuum, $B - V$ reddening, and H$\alpha$ emission
occurred, which was accompanied by 
increased absorption in the soft X-ray region (SLM). In contrast, 
the near disappearance of $f_{82}$ {\em preceded} the 2010 outburst
by some five years and therefore was unrelated to that event. Thus,
contra LB21, it is not clear that this event was associated with changes
in strength of dominant nonradial pulsation (NRP) modes.

\subsubsection{Rapid amplitude changes of coherent signals}
\label{suddn}

The full 
amplitudes listed in Table\,3 are values averaged over several seasons. 
From our short intensive-night campaigns (up to 7${\frac 12}$
hrs per night), we found that the amplitudes can vary unexpectedly.
An important result coming out of our intensive monitoring
on short order
is that the normally dominant signal at $f_{82}$ was sometimes
eclipsed by another, nominally secondary, signal.
Table\,4 summarizes the results of several intensive mini-campaigns 
and the secondary frequencies (full amplitudes) they exhibited.  
Because we sometimes found temporarily dominant frequencies during
these brief campaigns, their short-term behaviors suggest that their
amplitudes vary much more frequently than we would infer from periodograms 
drawn from a full season or longer.  Perhaps this ``flutter" of signals 
is typical.

\begin{table}
\begin{center}
\caption{Results summary of intensively observed nights }
\begin{tabular}{@{}lcrrl@{}}
\hline \hline
Season &  \# Nights & Freq. & Ampl.  &  Comments \\
 2000  &  4  &   5.03  &    6  &   fits last 2 nights \\
 2001  &  2  &   0.82  &    7  &   \\
 2003  &  6  &   0.82  &   11  &  waveform change  \\
 2004  &  3  &  1.24(?),~0.82 &  12 &   evolving freqs. \\
2011  &  1  &   --      &  $<$\,2 &  flat over 5 hrs \\
2016  &  2  &   2.48   &   18 &  \\
\hline
\end{tabular}
\end{center}
\end{table} 

We discuss the results from our short dedicated-night campaigns season by 
season as follows: \\

\noindent Season 2000:
This season included two pairs of intensive monitorings (each separated by
2-3 nights) 
spaced a month apart. Taking the second pair first, 
their variations could be fit with a signal of $f$ = 5.03\,d$^{-1}$ 
and a mean full-amplitude of 6 mmags. The behavior of the data in the 
first pair was decidedly different, exhibiting only small variations 
during their 4-4${\frac 12}$ hr coverages.

The $B, V$ periodograms for the whole Season 2004 exhibited peaks of 
amplitude 4.8-5 mmag at 1.243\,d$^{-1}$, 
as well as a strong transient signal 
(amplitudes 5-7\,mmag) at a frequency of 0.76 d$^{-1}$, in addition 
to the neighboring f$_{82}$. Though apparently real, this transient
did not recur for any other season, and so we have not listed it 
in our tables as a multiseasonal signal.

\noindent Season 2001: The $f_{82}$ full amplitude  over two
consecutive nights, 7\,mmag, is typical for the season (Table\,2).

\noindent  Season 2003: Amplitudes of all three secondary frequencies
were low or invisible during this season, signifying that the waveform changes
discussed in $\S$\ref{skwns} are unlikely to be due to intermode beating.

\noindent  Season 2004: Light curves for a 
sequence of three consecutive nights were conspicuous with f$_{82}$
appearing to beat with a sinusoid consistent with $\approx$1.2\,d$^{-1}$. 
The combined full amplitude was large (12\,mmag).

\noindent  Season 2011:
No variations were found over 5\,hrs.

\noindent  Season 2016: 
The observations of these two consecutive nights are the only ones
observed when the 2.48\,d$^{-1}$ signal was dominant.  Its amplitude 
then is among the largest found during all our monitoring of \gc. \\

We temper these descriptions by noting that our fittings of large-amplitude
sinusoids to data of only several hours of a few nights cannot be 
differentiated in general from beating by roughly similar 
modal amplitudes. The best single case for an amplitude waxing and waning 
within 1--2 weeks is discussed just below.

To summarize, from the well-observed nights referred to in Table\,4, one 
sees frequent changes in amplitudes not only for $f_{82}$ but also for  the 
``secondary" signals near  1.24\,d$^{-1}$, 2.48\,d$^{-1}$, and 5.03\,d$^{-1}$,
all probably occurring on rapid timescales. 
It is not surprising to find in \gc~what are evidently NRP modes excited 
in this frequency range, as 
it is a star situated at the hot edge of the $\beta$\,Cep domain. 
Although the $f_{82}$ amplitude starts to decrease in 2005--2012, 
and is generally mimicked by f$_{1.24}$ and f$_{5.03}$, the amplitude
of f$_{2.48}$ increases during later seasons (Table\,3.)

\subsection{Searches for other frequencies}
\label{hghfrq}

High frequencies ($>$ 8\,d$^{-1}$) are of interest to this study because 
of their potential identification with NRP p-modes and in turn as a possible
cause of the {\it msf} in line profiles 
\citep[][``N20b'']{Naze et al.2020b}. However, the quality of our
APT periodograms deteriorates above 8--10\,d$^{-1}$ and is 
meaningless beyond it. However, N20c's published TESS periodograms 
for Sectors 17 \& 18 exhibit no signal in the range above 8 d$^{-1}$ 
out to 20 d$^{-1}$,
(just as LB21's periodograms show no high-frequency signal down to 
$<$0.1\,mmag, not only for this time period but also during Sector 24).

\subsection{Changes in $f_{82}$ waveform }
\label{skw}

\subsubsection {Rationale for analysis }
In Paper\,1 we discovered an unusual skewness in the mean waveform of 
the f$_{82}$ signal taken from the first several APT seasons.
For convenience we characterized departures from a sinusoid by  
parameters $e$ and $\omega$ taken from the familiar
Lehmann-Filhes equation for orbital solutions of radial velocity variations.
Here $e$ and $\omega$ are a fake ``eccentricity" and ``longitude of periastron,"
respectively. Parameter $e$ represents the waveform's pointiness while 
$\omega$ quantifies its skewness. Values in the fourth quadrant signify 
a depressed positive-phase wing; the first quadrant gives the opposite 
skewness.  In our implementation of past and current work, we used a 
generalized least-squares algorithm by \citet[][]{Markwardt2011} adapted 
for our computations.  For the first eight seasons (Paper\,1), the
resulting means, averaged over $B$ and $V$ filter datasets, were
$e$ = 0.35 and $\omega$ = +285$^{\circ}$.  In Paper\,2 we examined data 
for six dedicated and consecutive nights of observations in 
Season 2003 and found for the $V$ filter that $e$ had increased to 
0.51${\pm 0.05}$ and the skewness had reversed sign to 
$\omega$=+25$^{\circ}$$\pm{6}$$^{\circ}$. As noted then, nearly identical
values and errors were found in our data by Dr. Fekel using an independent 
algorithm. 
The departures from a sinusoid is a remarkable result and thus requires 
confirmation.  We note for completeness that the periodogram 
for these six nights' data exhibits a faint second harmonic feature.

To check the statistics in a different way, we conducted an experiment
adopting a simple  Monte Carlo strategy for the $V$-band dataset of another
observing season, 2001, and compared errors derived for  $e$ and $\omega$ from
fake datasets using the 2001 observation times and photometric errors. We
then  compared them with results from a direct analysis of Season 2003 data. 
Note that an assessment of fluctuations of both seasons's target 
and check-minus-comparison star data (outliers removed), revealed
no significant departures from Gaussian distributions. 

Our experiment began by phase-folding the Season 2001 $V$-band data to the
f$_{82}$ ephemeris. The best (though mediocre) sinusoidal fit was computed 
for these initial data, and this sinusoid was subtracted, resulting in an
initial data fluctuation array. We then conducted mock data simulations for
11 independent trials by shifting the fluctuation array by phase increments 
of N$\times$28 points according to each point's former position. Here N =
0, 1, 2,...,10 corresponds to trial number and 28 is an arbitrary value,
which brings each point to a different observation time and night.
For each trial, fluxes of the so-shifted 
fluctuation array were added back to the initial fitted sinusoid,
and $e$ and $\omega$ were solved for again. 
Finally, their means and r.m.s. errors from these trials were computed
and compared with the original $e$ and $\omega$ computed for Season\,2001.
The mean values turned out to be nearly coincident with the solution for 
the original Season\,2001 data, namely means of $e$ = 0.16 and 
$\omega$=310$^{\circ}$ and r.m.s. errors of
$\sigma$$_e$$= {\pm 0.077}$ and $\sigma$$_{\omega}$ = ${\pm 8}$$^{\circ}$. 

We repeated this exercise by sampling only every other 
observation. The new r.m.s. values were $\sigma$$_{e}$ = ${\pm 0.102}$ 
and $\sigma$$_{\omega}$ = ${\pm 12}$$^{\circ}$ and thus scaled
approximately with the inverse square root of the number of points. 
Applying the same scaling for like-quality and increased 
numbers of observations to  Season 2003 (Table\,2),
we were able to predict errors of $\sigma$$_{e}$ = ${\pm 0.055}$ and 
$\sigma$$_{\omega}$ = ${\pm 6}$$^{\circ}$. 
These error estimates are nearly the same as Paper\,2's results from a direct
analysis of Season 2003 (viz., ${\pm 0.05}$ and ${\pm 6}$$^{\circ}$). We will
now use this result in the foregoing analysis of data subsets of this season.

\begin{table}
\begin{center}
\caption{Lehmann-Filhes waveform parameters, Season 2003}
`\begin{tabular}{@{}l|ccc|ccc@{}}
\hline \hline
  Filter & Halves 1+2  & (All but  &  6 nts.) & Intensive &  (Only & 6 nts.) \\
       & Ampl. &  $e$    & $\omega$ & Ampl.  & $e$  & $\omega$ \\
  B    & 5.2  & 0.41    & 348$^o$  & 12.1 &  0.50  & +20$^o$ \\
  V    & 4.2  & 0.41 &  307$^o$ & 10.2 &  0.47 &  +19$^o$ \\
       &      &      &    &   &    &    \\     
\hline
       &   Half 1 &  &   &          Half 2  &   &   \\
  &   Ampl. &  $e$ &   $\omega$  & Ampl. &  $e$  &  $\omega$ \\
  B    & 5.5  & 0.34 & 350$^o$ & 5.7 &  0.43  & 340$^o$ \\
  V    & 5.1  & 0.39 & 340$^o$ & 5.1 &  0.41  & 319$^o$ \\
\hline
  \end{tabular}
\end{center}
\noindent Note: 
The set of six ``intensively" monitored nights occurred between 
two nearly equal time segments during this observing season.
\end{table}

\subsubsection {\normalfont{Season 2003 waveform changes}}
\label{skwns}

We proceeded to analyze the waveform for $B$ as well as $V$-band observations 
of our six intensively monitored nights from Season\,2003 (Reduced Julian 
Day\,52,962--52,967). We will then contrast it to 
observations taken from  the rest of the season.  
In comparison to Paper\,2's analysis, we incorporated differences in outliers
comprising  the seasonal dataset and  used
different procedures for prewhitening of the long period in this analysis.
We remind the reader that we had found amplitudes from the secondary signals
to be low or absent in this season.

To examine the waveform differences during this season, we 
used the fact that the six intensively monitored nights occurred in
the middle of the observing season. We divided the datasets for the 
non-intensively observed nights into two halves and compared the 
resulting waveforms of all four groups -- viz.
first-half, second-half, both halves (all nonintensive nights), and the six
intensive nights. 
The $e$, $\omega$ parameter determinations for these groups are given in 
Table\,5. The most obvious result is that the full amplitude doubled from 
5--6 mmags to $\sim$11 mmag, and then decreased to its former value. 
Also, the ``eccentricity" for the six nights increased to 0.50 and 0.47 
for $B$ and $V$, respectively. 
This increase in $e$, +0.16, is more than double the
predicted $\sigma$$_{e}$ of ${\pm 0.055}$, according to our control 
results for Season\,2001 (when scaled for numbers of data points). 

These differences are visible in the phase-folded plot, Fig.\,6. 
Here the Lehmann-Filhes fit to the six-night, $B$-filter data points 
is displayed as a solid line. We can contrast it
with the Lehmann-Filhes fit to the data for all other nights in the
season (dashed line) and also with the best, though mediocre, sinusoidal
fit to the data for these other nights (dot-dot-dashed line).
The differences between the values of $e$ and $\omega$ and the 
published values in Paper\,2 are comparable to the error 
estimates found in our control: 
 $e$ = 0.50 here vs. 0.55 in Paper\,2, and $\omega$ = +19$^{\circ}$ here
vs. 25$^{\circ}$ there.
Also, just as with the eccentricity, its skewness 
subsequently reverted to its typical fourth-quadrant sense. 

\begin{figure} 
\begin{center}
\includegraphics*[width=6.2cm,angle=90]{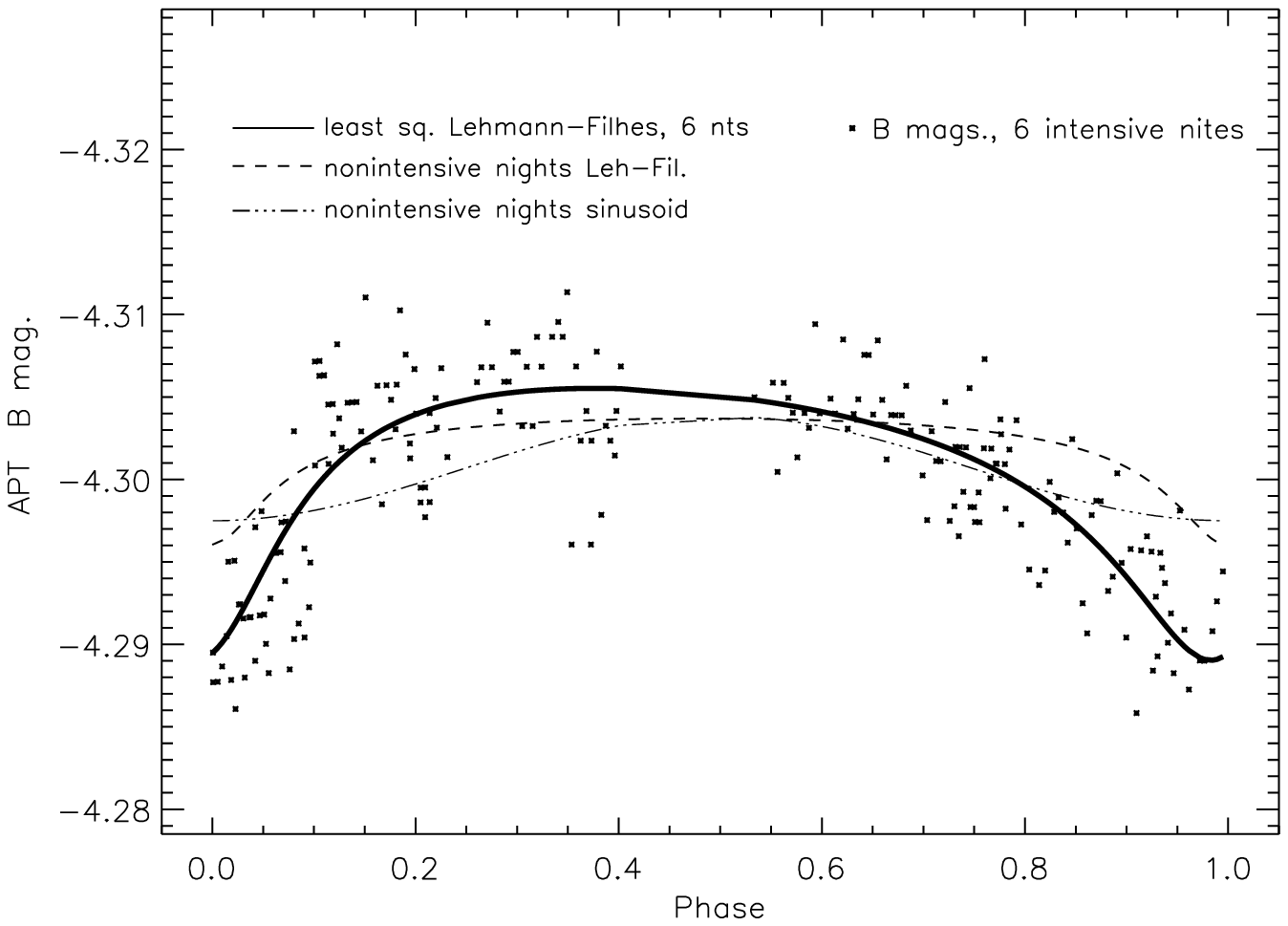}
\end{center}
\caption{The waveform determined by least-squares fit to the Lehmann-Filhes
solution for the $B$-filter, f$_{82}$ phase-folded data for six 
intensively monitored nights in Season 2003 (heavy curve and dots). 
Note the departure of the data and this curve 
from the sinusoidal fit (dot-dot-dashes) and the Lehmann-Filhes 
fit (dashes) for the nonintensive nights: the 6-night
solution has a larger amplitude and exhibits a ``pointy/bowed" form and 
skewness with a steep negative-phase wing. 
 }
\end{figure}

\subsection{The UV continuum dips}
\label{uvcuvc}

\begin{figure} 
\label{uvcs}
\begin{center}
\includegraphics*[width=6.2cm,angle=90]{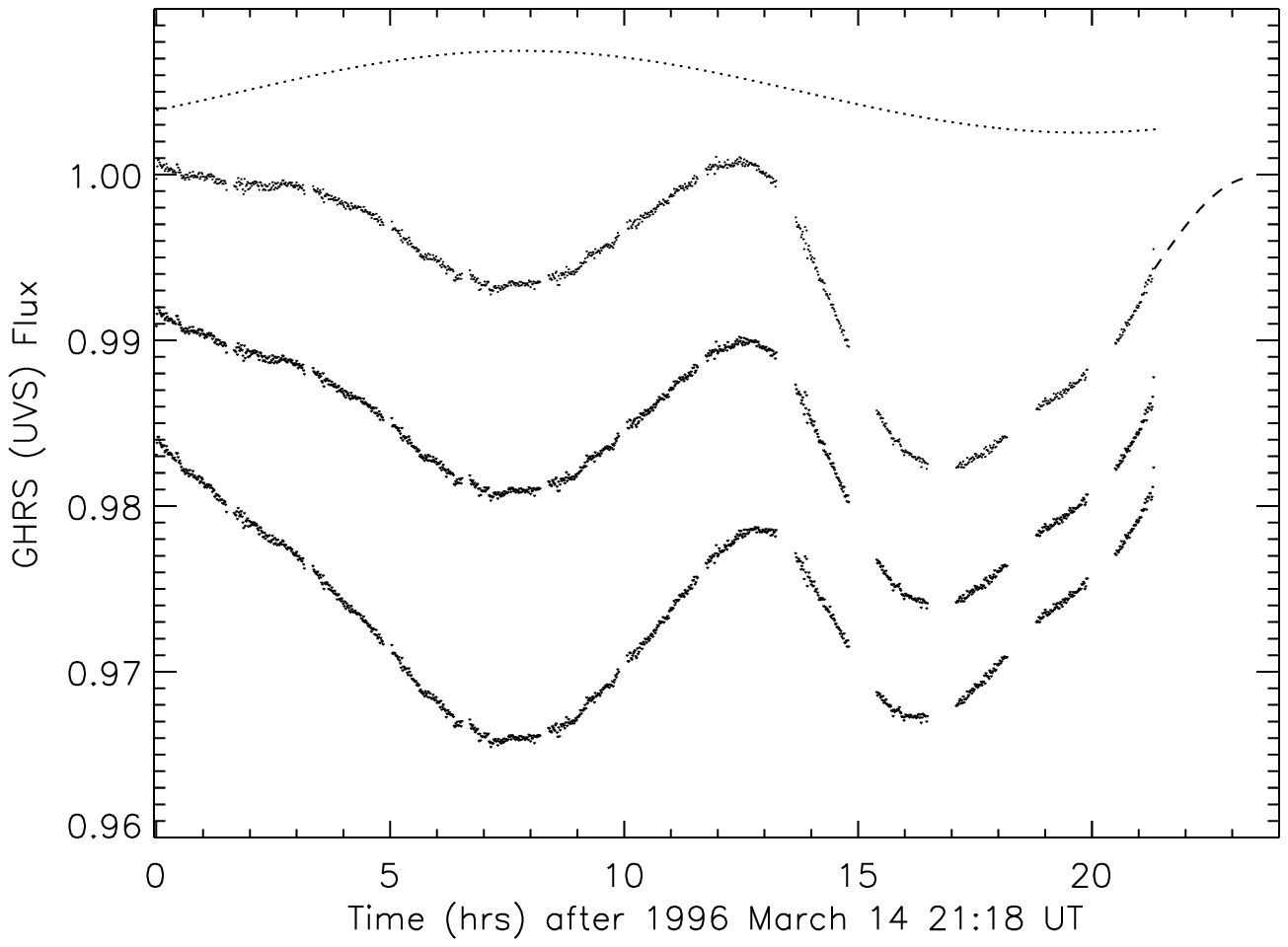}
\end{center}
\caption{The GHRS UV continuum light curve of 1996 March 15
(upper full curve).  The top line (dotted
sinusoid) is the $f_{82}$ signal from its ephemeris of Table\,3. 
The second and third curves are the original GHRS curve with the top
sinusoid subtracted by a factor of one and 2.5, respectively. 
One or the other of the lower curves shows how the undistorted UVC should 
appear if no $f_{82}$ signal existed.
}
\end{figure} 

Although the two IUE UV light curves of \gc\,lasted longer, the GHRS  series
of 1996 March 14-15 is unmatched in its precision. In Fig.\,7 we have 
exploited this fact and the f$_{82}$ ephemeris to represent this signal
as a 6 mmag sinusoid (dotted line) against the GHRS data (first full curve).
The error in the phase-positioning of the sinusoid is $\pm{0.01}$ cycles. 
We have subtracted the sinusoid from it to show how it would look if 
$f_{82}$ were not present (second full curve). This would be appropriate, 
for example, if f$_{82}$ is a very low-frequency, ``classical" NRP g-mode, 
for which geometric distortion of the star would dominate flux variations. 
The amplitude of the variations would then be wavelength-independent.
If instead, and as argued by SRH, the absorptions arise from a cool 
intervening cloud, then the wavelength dependence grows in the far-UV, and 
the GHRS light curve will approximate the lower curve. In any case, the lower 
curve gives an approximation of a sinusoid, for which 
least-squares fitting gives a dip separation of 8.90${\pm 0.02}$ hr 
(2.70${\pm 0.01}$\,d$^{-1}$).
However, it is easy to show from the 1982 and 1996 IUE records,
and even the GHRS curve, that these variations are not part of a 
true sinusoidal signal. Even so, S19 pointed out that the 43\,hr-long IUE
sequence suggested the presence of a third dip that seems to be the 
recurrence of the first dip from one rotation cycle earlier.

\begin{figure} 
\label{frq07}
\begin{center}
\includegraphics*[height=7.2cm,width=0.35\textwidth,angle=90]{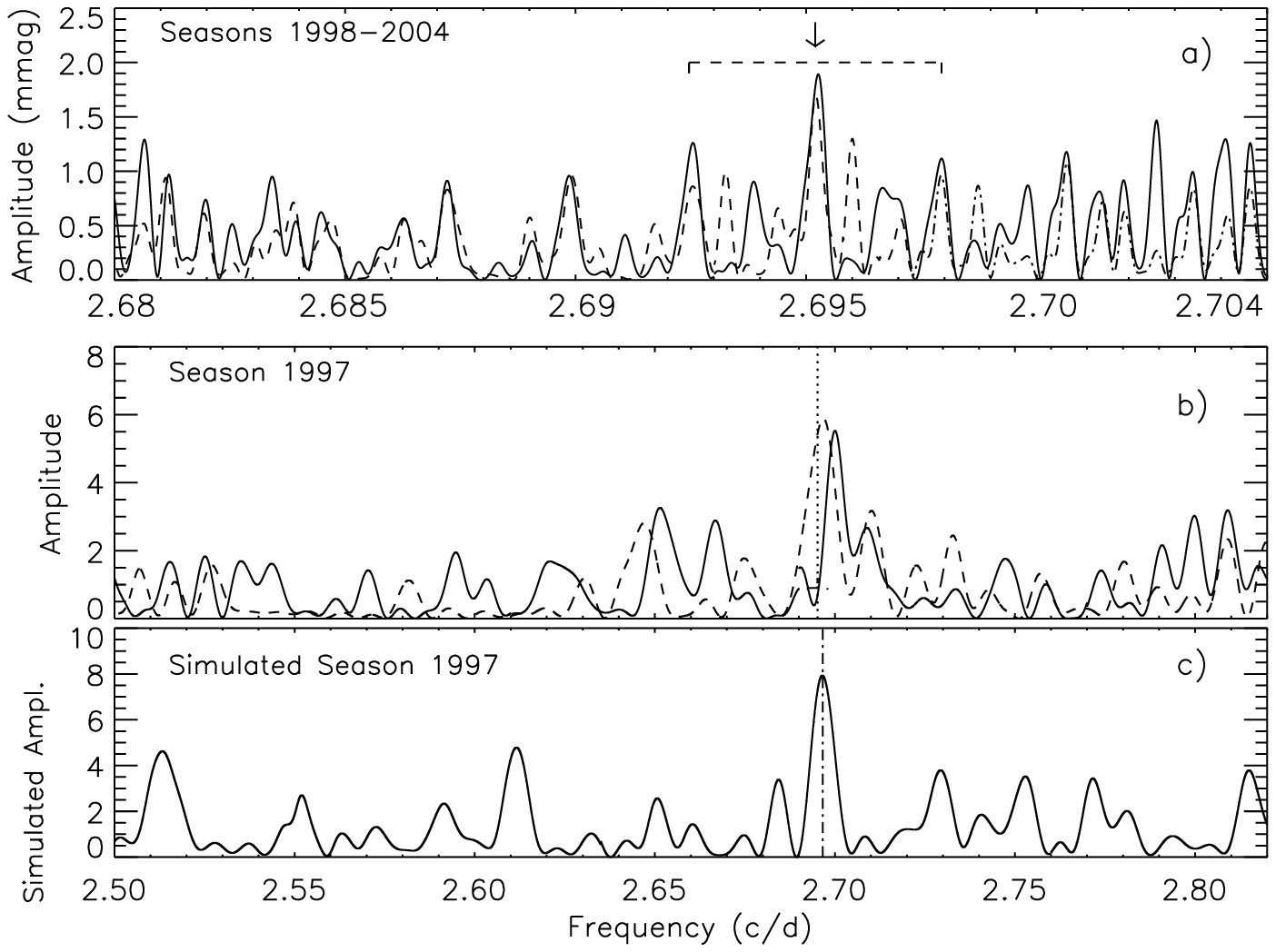}
\end{center}
\caption{Periodograms for $B$ (dashed line) and $V$ (solid line) covering 
a weak signal at the expected frequency 2.695\,d$^{-1}$. 
Panel a): results for Seasons 1998--2003. The horizontal line and
markers indicate the feature's annual sidelobes.
Panel b): the same feature for just Season 1997. Panel c): the simulation of
the GHRS signal applied to Season 1997 (see text).  For Panels b) \& c) we 
extend the frequency range to 2.5-2.82\,d$^{-1}$ to show the noise level. 
(The slight positional differences of the signal is not significant.)
The dotted line in Panel b) is the extension of the signal arrow in 
Panel a).  This position is almost identical to the dot-dashed line (Panel c).}
\end{figure}

One can ask whether the 8.9-hr separated dips are coherent enough through
the years to be detected in the APT data, since the program started just
over one year after the GHRS campaign. Yet even if it was present, it is not 
a continuously repeating sinusoidal signal. An intermittently occurring dip
will cause the periodogram to show a more complicated beating structure.
Fortunately, because we now know where to isolate a narrow search range, 
we can search a periodogram for a peak that emerges at a predicted 
frequency of 2.70\,d$^{-1}$. We first searched for signals in the $B$ 
and $V$ periodograms in the 23-season composite and found none. 
However, we did find a possible weak peak at 2.695\,d$^{-1}$ in the 1998--2004
multiseason periodograms, which are shown in Fig.\,8a. 
Their annual sidelodes are also visible. To see if this candidate signal also 
exists for Season 1997, the season closest to the UV 1996 campaigns, 
we computed the periodograms for this season and display them in Fig.\,8b to 
show that they have the same peak.  Note that although the Season 1997 noise 
is twice as large as Season 1998-2004, the signal is three times stronger.
From these results, the putative signal is weakening and beomes invisible
in late APT seasons. Although there is no immediate way of checking,
it is possible that in recent years the UV dips have no longer been present.

To verify our Season 1997 detection, we have constructed artificial 
datasets by estimating the egress wing to the
pre-dip level and thus completing the expected UVC curve 
out to 1.216\,days from observation start (see 
dashed line in Fig.\,7). We then repeated and concatenated the curve 
115 times until it covered the full span of this season.  
Since the signal-to-noise of the GHRS greatly exceeded the APT's, we 
then added Gaussian noise to simulate the APT's data 
fluctuations. Finally, we sampled the noisy, season-long curve at
the actual observation times. We repeated the procedure for various assumed
noise levels. The result was a series of mock light curves for 
various noise levels that retain the observing window gaps. 

Figure\,8c shows the resulting periodogram for a mock SNR of 250 per 
observation, which matches the measured APT r.m.s. 
The position of the generated signal, at 
2.697\,d$^{-1}$, is already ``locked in" to the 8.90\,hr we measured earlier, 
so its position is a given. Similarly, the SNR of the 
peak height to the surrounding noise level is not unexpected either because
the estimated noise level of the observations was determined from the
scatter of comparison-check star observations (see $\S$\ref{obbs}).  
However, the near agreement of the simulated and observed peak heights in 
Figs.\,8b \& 8c reassures us that the signal is stellar. Because the strengths 
of the dips are already known to vary over time (e.g., the dips were 
stronger in 1982), we consider this amplitude agreement fortuitous.

\section{Discussion}
\label{dscss}

\subsection{High frequency pulsations and migrating subfeatures}
\label{hifrq}

High-frequency modes are now known to be active in Be stars, e.g., 
in $\pi$\,Aqr, a \gc\,analog, which exhibits one or more 
tesseral modes \citep[][``N20b'']{Naze et al.2020b}. 
The amplitudes of these features traveling through line profiles amount to
several percent  in line profiles  (1--2.5\,mmag in the optical light curve).  
In N20b's Fig.\,4 one sees that the spacings of the NRP-induced migrating 
subfeatures in optical spectra of $\pi$\,Aqr, 
undoubtedly due to high-frequency p-modes, are rather uniformly spaced 
in time.  Thus, there are more differences than similarities in 
the {\it msf} of this star's spectrum compared to  \gc.

In view of the  results on $\pi$\,Aqr, we now revisit 
the \citet[][]{Yang et al.1988} and our own reports of
migrating subfeatures in optical and UV spectral profiles of \gc.~
In this context, L20 have doubted our interpretation of these features as 
clouds forced into corotation over transient magnetic centers on the star
\citep[][]{Smith&Robinson1999}. Therefore, we now address arguments
for an NRP interpretation for the {\it msf.}

The most precise record of migrating subfeatures in the spectrum of \gc\,is 
the 21.5\,hr time series of GHRS UV spectra.  In this series, a 
raft of {\it msf} associated with many lines is ubiquitous.
The features traveled across line profiles at a rate of 
+95${\pm 5}$\,km\,s$^{-1}$\,d$^{-1}$ and reoccur at unpredictable intervals 
averaging very roughly 2 hours (12\,d$^{-1}$) during this time series. 
Importantly, they wax and wane in visibility 
in about 1$\frac{1}{2}$\,hrs. In a separate study
\citet[][]{Smith1995} discussed 109 high-dispersion difference spectra of
the He\,I $\lambda$6678 profile during five nights in 1993; each monitoring 
interval was 3--6 hours.  Difference spectra revealed 
{\it msf} striation patterns that reoccurred at erratic intervals on most
nights and lasted no more than 2-3 hours. The acceleration rate of the
features was +92${\pm 10}$\,km\,s$^{-1}$\,d$^{-1}$, in good agreement with 
the later UV results. 
Similarly, the unpredictable appearances and short lifetimes of these 
events render any attempt to measure their recurrences all but meaningless.
The cyclical intervals between these patterns averaged 2--2$\frac{1}{2}$ hrs 
in the 1993 monitoring and 1$\frac{1}{2}$--2 hrs for the 1996 UV monitoring.  
For the following discussion we note that the observed amplitudes 
of the {\it msf} were about 0.4\% for the optical He\,I line and (depending 
on the line's excitation) 0.3--0.6\% for the UV, i.e., 
the {\it msf} amplitudes are similar in the two wavelength regimes.

With this description, we discuss why high-frequency NRPs are not
the best explanation for the {\it msf} in \gc:~

\begin{enumerate}

\item The absorption features are noncoherent.  Also, unlike
NRP bumps in line profiles, they
are not necessarily most prominent in the middle of their lifetimes.

\item To match roughly the irregular spacings of the {\it msf} in the
GHRS dataset, an NRP p-mode would have a frequency of 9--12\,d$^{-1}$.
The reported signal at 7.57\,d$^{-1}$ (N20c, LB21) is too low to meet 
this criterion. 

\item  Consider that the
ratio of line profile-to-photometric {\it msf} semiamplitudes in 
${\pi}$\,Aqr is 2.5\% to 1\,mmag. These variations are due to NRP
\citep[][]{Naze et al.2020b}. Because the  
amplitudes of {\it msf} in \gc~spectra are five times 
smaller, the photometric amplitudes in \gc~periodograms should be 0.2 mmag 
if they too are caused by NRP. Since the TESS periodogram of \gc~rules out 
photometric amplitudes down to less than 0.1\,mmag at high frequencies,  
the line profile {\it msf} of \gc~are not likely to be caused by NRP. 

\item 
Within measurement errors, the amplitudes of {\it msf} are the same 
for optical and UV spectra. This is an important point because the restoring
force of high-frequency pulsations is due to pressure
imbalances (from their high frequencies they are likely to be p-modes). 
Model atmosphere simulations indicate that temperature variations from
pulsations represented by spherical harmonics cause  flux variations
in early-type B stars that are $\approx$2${\frac 12}$ 
times as large as in the optical, not equal to them as observed.

\end{enumerate}

Of these arguments, the third one is probably the most powerful.
However, the stipulation should be made that the amplitude of the 
line-profile and photometric {\it msf} reflected behaviors at different times.
The fourth argument relies on the wavelength-dependence of amplitudes
in the UV being large compared to optical for high-degree p-modes,
as is true for low-degree ones. At least for rapidly rotating, early-Be
stars this is relatively unexplored territory, and further exploration
is necessary. 
To date, the amplitude behavior with wavelength seems to have been 
investigated theoretically so far only up to 
intermediate-degree ($\mathit{l}$=3-4), classical g and p modes 
\citep[e.g.,][]{Pigulski et al.2017}.

We remark further that if UV-absorbing structures are suspended 
over different stellar latitudes, their signatures will exhibit more than 
one acceleration across the line profiles.

\subsection{The nature of the 0.82\,d$^{-1}$ frequency}


As part of our initial justification for assigning $f_{82}$ to rotational 
modulation, Paper\,1 argued that the alternative, classical NRP
modes were not likely to be excited in rapidly rotating
early-type Be stars, whereas this frequency is quite consistent with rotational
modulation of a surface inhomogeneity. As described next, this was likely 
to be a premature conclusion. In the meantime, various spectroscopic 
campaigns as well as a flood of results
and satellite photometric surveys 
have shown that nonradial pulsations are endemic to Be stars, including
early-type and rapidly rotating stars. 

NRP was given a
major boost from the study of spectral line profile variations for a
small but representative population of other early-type classical Be stars
\citep[e.g.,][]{Rivinius et al.2003}. More recently, the discovery of groups
of frequencies sometimes close to the rotation frequency, 
$\Omega$$_R$, has presented evidence that nearly all of them must be due 
to NRP  \citep[][]{Baade et al.2016}.  Even so, some investigators 
\citep[e.g.,][Balona 1995]{Balona&Engelbrecht1986} have argued that
one of these frequencies, particularly if it is visible at intermittent
intervals, ean be caused by rotational modulation of a starspot.

Satellite photometry has demonstrated that NRP modes are endemic to early-type 
Be stars.  Recent photometric satellite surveys 
\citep[e.g.,][``BO20a,  BO20b"]{Balona et al.2015, Semaan et al.2013, Semaan 
et al.2018,Saio et al.2017,Labadie-Bartz et al.2020,Balona&Ozuyar2020a,
Balona&Ozuyar2020b} have disclosed, for most stars exciting low-frequency
signals near their rotation rates ($\Omega$$_R$), 
that these signals are members of clusters of frequencies, wherein only 
one at most can be rotational.

 Yet, one can ask whether there exist rapidly rotating, early-type Be stars 
with single isolated frequencies near $\Omega$$_R$?
\citet[][]{Balona2020} reports that in the larger TESS survey 
of BO20b, five O9--B2 stars 
exhibit apparent isolated, coherent modes at frequencies that arguably
coincide with the rotation frequency.
Therefore, because such signals that meet our conditions do exist in a few 
early-type Be stars,
it is now apparent we can no longer hold, as in our 
previous papers, that the isolated signal at $f_{82}$, though arguably close 
to $\Omega$$_R$, is unique to \gc.



Before interpreting the $f_{82}$ frequency as arising from a long-period
g-mode, one should consider a potential obstacle. This is that 
an oscillation observed near $f_{rot}$ in the inertial frame 
will have a frequency of nearly zero in the corotating frame, i.e., the 
period will be very long in the (physically important) reference frame.  
Its corresponding peak in the periodogram would blend with 
its high-order neighbors, causing a broad peak, which
is not observed. A better identification would be of 
an r-mode, which is an essentially horiontal vorticial
pattern at the surface excited by Coriolis force imbalances (or the 
$\kappa$ mechanisam \citet[][]{Saio et al.2017}), to which we turn next.
\


Several years ago \citet[][]{Walker et al.2005} reported the
excitation of a thicket of low frequencies ($\ltsim$0.005 mHz) in the rapidly 
rotating  B5e star HD\,163868. Unlike several other modes of frequencies
0.02  mHz or higher, which can be ascribed to $p$ or $g$ modes, this
low-frequency cluster is close to a multiple of the rotational frquency 
(i.e., 0.90-0.95m$\Omega$, where perhaps $m$ = 1). According to
\citet[][2018]{Saio2013}, these are probably signatures 
of odd, low azimuthal order r-modes. 
The circulation of surface particles participating in r-modes are associated
with large polar-direct velocity components at mid-latitude. 
This renders the discovery of r-modes particularly advantageous in stars 
observed at intermediate inclination like HD\,163868.

Since \gc~is likewise observed at an intermediate inclination, its $f_{82}$
signal may likewise be due to a low-degree r-mode. 
Notably, r-modes are predicted to occur as clusters of frequencies, often
adjacent to a dominant one \citep[e.g.,][Saio et al. 2018]{Walker et al.2005}.
From previous and present APT results, 
it has appeared to us thus far as if only an isolated peak is visible 
near $f_{82}$. If  $f_{82}$ is indeed an r-mode, other associated 
r-modes may be present with amplitudes too small to be detected by the APT.
Thus, LB21's detection of multiple low-amplitude g$_1$ modes 
in their TESS light curve appears to be consistent after all with f$_{82}$ 
being the dominant mode of an r-mode complex, at least when it was
visible in early years of the APT program. In addition to \gc~and 
HD\,163868, a few rapidly-rotating B stars in the cluster NGC\,3766 may 
well excite both g- and r-modes \citep{Saio et al.2017}.

\subsection{Extra-APT contributions}

 As few in number as the UV satellite monitorings of \gc\,are, combined with
the revised ephemeris of Table\,3, they permit a reinvestigation and 
an entry of new evidence as to the origin of the $f_{82}$ signal. 

We measured the times corresponding to passages of the centroids of the 
{\em first} dip in the IUE 1982, IUE 1986, and GHRS light curve (Fig.\,7). 
These occur respectively at RJD's 44997.48, 
50101.57, and 50157.70. According to our Table\,3 ephemeris, these times 
correspond to faint-star phases 0.41, 0.92, and 0.08, respectively. 
We estimate errors on the 1996 IUE and GHRS phases as ${\pm 0.02}$ and 
${\pm 0.03}$. These values are dominated by errors 
in our adopted frequency (Table\,3). For the more important
{\em phase-difference} error between the two 1996 first-dip centroids (the
time interval  being 57 days), the error in the frequency is negligible. 
The error in the 1996 IUE feature relative to the GHRS feature is dominated 
by centroid measurement and is ${\pm 0.02}$.
The errors for the 1982 IUE dip are larger, ${\pm 0.05}$, because the 
frequency and centroid-finding errors must be folded together. 
Similarly, phases for the centroids of the {\em second} UV dip 
are 0.71, 0.22, and 0.39. We estimate phase errors in the GHRS second 
dip to be dominated by uncertainty in frequency, whereas errors in the IUE 
second-dip centroids are about 50\% larger because their profiles are not
as well defined. 
If we had instead adopted the B20 frequency and our T$_{\circ}$ from Table\,3, 
the phases would be different, but the net result is much the same. 
All told, the far-UV features are not phase-locked with the 
optical ephemeris.
The phasing mismatch occurs even over the small interval (46.16 cycles) 
in the $\approx$57 day interval between the 1996 IUE and GHRS observations.
Assuming a cycle miscount of 0 or 1, the percentage mismatch would be either
0.4\% or 2.5\%, respectively, which, though small, introduces a phase
slippage.

The resolution of this slippage starts with the fact that only 
one frequency can be rotational.
We believe it is unrealistic that, over the 57 day interval between 1996 IUE
and GHRS observations, a surface differential rotation rate
of order 1\% or more occurs. 
Therefore, we reject this possibility. It follows that we {\em prefer} 
to adopt the alternative: $f_{82}$ is not a rotational signal. Yet, our 
preference does not prove the case. In fact, given the isolation
of the signal at low frequency and its sometime nonsinusoidal waveform, the
identification of f$_{82}$ with NRP is not straightforward. In any case, in
view of the arguments put forth regarding the UV color changes of the dips 
(SRH) and the correlation of appearances with changes in UV 
spectral lines and hard X-ray flux (SLM16),
we can see no reason to reject the co-rotating picture. This does not mean
that the rotation frequency has been found.  Given the physical
parameters of the star and the likely reoccurrence of the ``first dip"
in the 1982 IUE light curve, it is probably near 0.8\,d$^{-1}$.

\section{Summary and Conclusions}
\label{smmry}

We summarize the main points of this work as follows:

\begin{itemize}
\item[$\ast$] 
The discovery of coherent signals with frequencies at 1.24, 2.48, 
and 5.03\,d$^{-1}$ found by B20 or N20c has been confirmed by an
independent dataset (APT).

\item[$\ast$] 
We agree with previous authors that these signals are NRP (p- or low-degree g-)
modes. These modes should not be confused with the stochastic low-frequency
($\sim$0.1\,--\,2\,d$^{-1}$) variability discovered in a variety of OB stars 
in TESS data \citep[e.g.,][]{Bowman et al.2020}.
For early B stars the
amplitude of such``white noise" is generally only $\sim$0.1 mmag in 
B\,III-V stars and thus is well below the detection limit of the APT. 
This variablity is thought to be excited by turbulence generated within the 
Fe-opacity convective zone \citep[e.g.,][]{Cantiello et al.2021}.

\item[$\ast$] 
The 0.82238\,d$^{-1}$ frequency 
 was stable in $B, V$ filters from 1997 through 2011, although the amplitude 
varied.  Since then, the signal has faded and then showed a weak recurrence 
(2014-2016). We cannot verify that it was active during 2017-2019.

\item[$\ast$] NRP amplitudes can at times wax and wane
rapidly. Such activity is also displayed in N20c's TESS dynamic periodogram
and Fig.\,9.

\item[$\ast$] We believe the 0.82\,d$^{-1}$ signal is an excited NRP 
mode of still undetermined type. However, its occasional tendency to modify
its waveform complicates a physical description of its origin.
Also, the isolated 
position of a detectable signal near $\Omega$$_R$  (but how near?) may not 
be unique among early Be stars (BO20b), but it is not the norm either.
Otherwise, given the excitation of these modes,
\gc~has begun to resemble other Be stars at the 
periphery of the $\beta$\,Cep domain.

\item[$\ast$] 
Our preference for attributing the cause of $f_{82}$ to NRP rather than  
rotational modulation was facilitated by using UV light curves. 
These indicate phase shifts from our $f_{82}$ ephemeris. 
We identify the pair of dips from UV photometry as being likely 
due to rotation. However, because we do not know the putative
stellar longitude separations of the absorbing structures, we cannot
determine the exact value of $\Omega$$_R$.

\item[$\ast$] 
In periodograms of early APT season data we found a signal at 2.70\,d$^{-1}$ 
that corresponds to the time separation of the two-dip pattern observed in 
the three UV campaigns of 1982 and 1996. 
However, periodograms of (most) later APT seasons and recent satellite 
datasets do not exhibit this signal. We conclude that it has diminished
and may no longer be visible in the UV or optical.

\item[$\ast$] 
TESS data offer no evidence that high-frequency NRP modes produce
the migrating subfeatures in optical and UV spectra of 
\gc.\,Moreover, the {\it msf} are chaotic, exhibit much larger amplitudes 
than the APT and TESS detection thresholds, and do not show an expected 
increase in amplitude from the optical to far-UV.


\item[$\ast$] The so-called long cycles abruptly faded to invisibility just
after the era covered by Paper\,2 and shortly after the 2010 outburst.
(A possible recovery in  two recent
seasons might have occurred, Fig.\,2, but is too weak to be reliable.)
We speculate that the continued build-up of the inner disk,
according to APT photometry and especially the increased 
He\,I $\lambda$6678 emission \citep[][]{Pollmann et al.2014}
is caused by an increased density there that overwhelms
a fragile disk dynamo mechanism.

\item[$\ast$] 
The correlation of APT and X-ray long cycles argues that the Be disk
mediates the production of hard X-rays on the star. 
Optical variations found by the APT have therefore been important
in framing the magnetic interaction hypothesis. 

\end{itemize}

In criticizing the star-disk magnetic interaction hypothesis, 
 L20 and B20 overlooked some key points.
The existence of the $f_{82}$ signal is largely irrelevant to the 
production of hard X-ray flux in this picture.\footnote{The rotation 
rate comes into the discussion of details of corotating clouds, and 
then only secondarily in estimating their elevations.  As such
the error in equating f$_{82}$ to $\Omega$$_R$ cannot be large.} 
The spectral {\it msf} are supportive though not essential to the basic
picture unless they can be identified with large \xr\,``flares."  
However, if instead they turn out to be due to high-frequency NRPs after all,
the case for suspended {\em cloudlets} would disappear. The optical/X-ray 
long-cycle connection {\em is} important to the picture -- the previous 
points are
not required.  The high densities associated with the flares strongly suggest 
a photospheric origin, to say nothing of the correlation of hard X-ray
fluxes with photospheric UV line strengths (SR99). Therefore, any rapid, 
aperiodic events, e.g., caused by emerging magnetic structures, should be 
examined as aiding in the understanding of the X-ray formation process.

\section*{Acknowledgments}
This study would not have been possible without the support of Lou Boyd, who
has conscientiously managed the automatic telescopes at 
Fairborn Observatory.
We are indebted to Drs. Yael Naze, Gregor Rauw, Andrzej Pigulski, and Mr. 
Piotr Kolaczek-Szymanski for sending us satellite light curve extractions
of \gc~and for many suggestions that improved the manuscript.  
We are also grateful to Drs. Raimundo Oliveira de Lopes and Christian Motch
for comments on the draft. 
It is a pleasure to acknowledge an anonymous referee's calling our attention 
to an important error in an earlier version of the manuscript and for 
many insightful suggestions.
GWH acknowledges long-term support from NASA, NSF, Tennessee State University,
and the State of Tennessee through its Centers of Excellence Program.



\begin{thebibliography}{}

\bibitem[\protect\citeauthoryear{Baade et al.}{2016}]{Baade et al.2016}
Baade, D., Rivinius, Th., Pigulski, A., et al. 2016, A \& A, 588, 56B

\bibitem[\protect\citeauthoryear{Baade et al.}{2018}]{Baade et al.2018}
Baade, D., Pigulski, A., Rivinius, A., et al. 2018, A \& A, 610, 70B

\bibitem[\protect\citeauthoryear{Balona}{1995}]{Balona1995}
 Balona, L. A., 1995, MNRAS, 277, 1547B

\bibitem[\protect\citeauthoryear{Balona}{2020}]{Balona2020}
Balona, L. A. 2020, priv. commun.

\bibitem[\protect\citeauthoryear{Balona et al.}{2015}]{Balona et al.2015}
 Balona, L. A., Baran, A. S., Dasy\'nska-Daszkiewicz 
et al. 2015, MNRAS, 451, 1445B

\bibitem[\protect\citeauthoryear{Balona \& Engelbrecht}{1986}]{Balona&Engelbrecht1986}
Balona, L. B., \& Engelbrecht, C. A. 1986, MNRAS, 219, 131B

\bibitem[\protect\citeauthoryear{Balona \& Ozuyar}{2020a}]{Balona&Ozuyar2020a}
Balona, L, B., \& Ozuyar, D. 2020, MNRAS, 493, 252B (BO20a)

\bibitem[\protect\citeauthoryear{Balona \& Ozuyar}{2020b}]{Balona&Ozuyar2020b}
Balona, L, B., \& Ozuyar, D. 2020, arXiv:2008.06288v1 (BO20b)

\bibitem[\protect\citeauthoryear{Barry et al.}{1984}]{Barry et al.1984}
Barry, D., Holberg, J. B., Forrester, W. T., et al. 1984, 281, 755B


\bibitem[\protect\citeauthoryear{Borre et al.}{2020}]{Borre et al.2020}
Borre, C. C., Baade, D., Pigulski, A., et al. 2020, A\&A, 635A, 140B (B20)

\bibitem[\protect\citeauthoryear{Bowman et al.}{2020}]{Bowman et al.2020}
Bowman, D. M., Burssens, S., Sim\'on-D\'iaz, et al2020, A\&A, 640, A36

\bibitem[\protect\citeauthoryear{Cantiello \& Braithwaite}{2011}]
{Cantiello&Braithwaite2011}
Cantiello, M., \& Braithwaite, J. 2011, A\&A, 540A, 140C 

\bibitem[\protect\citeauthoryear{Cantiello  et al.}{2021}]{Cantiello et al.2021}
Cantiello, M., Lecoanet, D., Jermyn, A. S. 2021, arXiv:2102.05670v2 

\bibitem[\protect\citeauthoryear{Doazan et al.}{1983}]{Doazan et al.1983}
Doazan, V., Franco, M., \& Sedmak, G. et al.1983, A \& A, 171, 180D

\bibitem[\protect\citeauthoryear{Eaton et al.}{2003}]{Eaton et al.2003}
Eaton, J. A., Henry, G. W., \& Fekel, F. C. 2003, ASSL, 288, 189E

\bibitem[\protect\citeauthoryear{Gies et al.}{1998}]{Gies et al.1998}
Gies, D. R., Bagnuolo, W. G., Ferrara, E. C., et al. 1998, APJ, 493, 440G

\bibitem[\protect\citeauthoryear{Hamaguchi et al.}{2016}]{Hamaguchi et al.2016}
Hamaguchi, K., Oskinova, L., Russell, C., et al. 2016, ApJ, 832, 140H 

\bibitem[\protect\citeauthoryear{Harmanec}{2002}]{Harmanec2002}
Harmanec, P. 2002, Exotic Stars, ed. Tout, C. \& Van Hamme, W.,
ASP Conf. Ser., 279, 221M

\bibitem[\protect\citeauthoryear{Harmanec et al.}{2000}]{Harmanec et al.2000}
Harmanec, P., Habuda, P., Stefl, S., et al. 2000, A \& A, 364, L85H

\bibitem[\protect\citeauthoryear{HEASARC Rosat Data Archive}{2020}]
{HEASARC Rosat Data Archive2020}
HEASARC Rosat\_Data\_Archive 2020, https://heasarc.gsfc.nasa.gov/docs/cgro/db-perl/W3Browse/w3table.pl

\bibitem[\protect\citeauthoryear{Henry}{1995a}]{Henry1995a}
Henry, G. W. 1995a, in ASP Conf. Ser. 79, Robotic Telescopes,
ed. G. W. Henry \& J. A. Eaton (San Francisco: ASP), 37H

\bibitem[\protect\citeauthoryear{Henry}{1995b}]{Henry1995b}
Henry, G. W. 1995b, in ASP Conf. Ser. 79, Robotic Telescopes:  Current
Capabilities, Present Developments, and Future Prospects for Automated
Astronomy, ed. G. W. Henry \& J. A. Eaton (San Francisco: ASP), 44H

\bibitem[\protect\citeauthoryear{Henry}{1999}]{Henry et al.1999}
Henry, G. W. 1999, PASP, 111, 845H

\bibitem[\protect\citeauthoryear{Henry \& Smith}{2012}]{Henry&Smith2012}
 Henry, G. W. \& Smith, M. A. 2012, ApJ, 760, 10H (Paper\,2)



\bibitem[\protect\citeauthoryear{Labadie-Bartz et al.}{2020}]{Labadie-Bartz et al.2020}
Labadie-Bartz, J., Carciofi, A. C., Henrique de Amorim, T., et al. 2020,
arXiv:2010.139055v1

\bibitem[\protect\citeauthoryear{Labadie-Bartz et al.}{2021}]{Labadie-Bartz et al.2021}
Labadie-Bartz, J., Baade, D., Carciofi, A. C. et al., MNRAS, 502, 242L, 2021 
(LB21)

\bibitem[\protect\citeauthoryear{Langer et al.}{2020}]{Langer et al.2020}
Langer, N., Baade, D., Bodensteiner, J., et al. 2020, A \& A., 633A, L40 (L20)

\bibitem[\protect\citeauthoryear{Mamajek}{2017a}]{Mamajek2017a}
Mamajek, E. 2017a, J. Double Star Obsns., 2, 264M

\bibitem[\protect\citeauthoryear{Mamajek}{2017}]{Mamajek2017b}
Mamajek, E. 2017b, priv.commun.

\bibitem[\protect\citeauthoryear{Markwardt}{2011}]{Markwardt2011}
Markwardt, C. 2011, Markwardt IDL Library, 
 https://pages.physics.wisc.edu/$\sim$craigm/idl/idl.html

\bibitem[\protect\citeauthoryear{Mason et al.}{1976}]{Mason et al.1976}
Mason, K. O., White, N. E. \& Sanford, P. W.\ 1976, Nature, 260, 690M

\bibitem[\protect\citeauthoryear{Motch et al.}{2015}]{Motch et al.2015}
Motch, C., Lopes de Oliveira, R., \& Smith, M. A. 2015, ApJ, 806, 177M 

\bibitem[\protect\citeauthoryear{Murakami et al.}{1986}]{Murakami et al.1986}
Murakami, T., Koyama, K., Inoue, H., et al. 1986, ApJ, 310, L31M


\bibitem[\protect\citeauthoryear{Naz\'e et al.}{2020a}]{Naze et al.2020a}
Naz\'e, Y., Motch, C., Rauw, G., et al. 2020a, MNRAS, 493, 2511N  (N20a)

\bibitem[\protect\citeauthoryear{Naz\'e et al.}{2020b}]{Naze et al.2020b}
Naz\'e, Y., Pigulski, A., , Rauw, G., et al. 2020b, MNRAS, 494, 958N  (N20b)

\bibitem[\protect\citeauthoryear{Naz\'e et al.}{2020c}]{Naze et al.2020c}
Naz\'e, Y., Rauw, G., \& Pigulski, A. 2020c, MNRAS. 498, 3171N (N20c)

\bibitem[\protect\citeauthoryear{Nemravov\'a et al.}{2012}]{Nemravova et 
al.2012}
Nemravov\'a, J., Harmanec, P., Koubska, P., et al. 2012, A\&A, 537, 59-69N

\bibitem[\protect\citeauthoryear{Peters et al.}{2008}]{Peters et al.2008}
Peters, G. J., Gies, D. R., Grundstrom, E. D., et al. 2008, 686, 1280P


\bibitem[\protect\citeauthoryear{Peters et al.}{2013}]{Peters2013}
Peters, G. J., Pewett, T. D., Gies, D. R., et al. 2013, ApJ, 765, 2P

\bibitem[\protect\citeauthoryear{Peters et al.}{2016}]{Peters2016}
Peters, G. J., Wang, L., Gies, D. R., et al. 2016, ApJ, 828, 47P


\bibitem[\protect\citeauthoryear{Pigulski et al.}{2017}]{Pigulski et al.2017}
Pigulski, A., Baran, A., Bzuowski, M., et al. 2017, Proc. Polish Astron. Soc.,
Vol 5 ed. K. Zwintz \& E Poretti, 76P


\bibitem[\protect\citeauthoryear{Pollmann et al.}{2014}]{Pollmann et al.2014}
Pollmann, E., Vollmann W., \& Henry, G. W. 2014, IAU Info. Bull. No. 6109, \#1

\bibitem[\protect\citeauthoryear{Pollmann}{2021}]{Pollmann2021}
 Pollmann, E., et al. 2021, JAAVSO, 49, No. 1

\bibitem[\protect\citeauthoryear{Postnov et al.}{2017}]{Postnov et al.2017}
Postnov, K., Oskinova, L., \& Torre\'jon, J. M. 2017, arXiv:2017.00336v1

\bibitem[\protect\citeauthoryear{Ricker et al.}{2015}]{Ricker et al.2015}
Ricker, G. R., Winn, J. N., Vanderspek, R., et al. 2015, J. Astr. Telescopes,
Instr. \& Syst, Vol. 1, id 014003

\bibitem[\protect\citeauthoryear{Rivinius}{2003}]{Rivinius et al.2003}
Rivinius, Th., Baade, D., \& Stefl, S. 2003, A \& A, 411, 229R 

\bibitem[\protect\citeauthoryear{Robinson \& Smith}{2000}]{Robinson&Smith2000}
Robinson, R. D., \& Smith, M. A. 2000, ApJ, 540, 474R (RS00)

\bibitem[\protect\citeauthoryear{Robinson et al.}{2002}]{Robinson et al.2002}
Robinson, R. D., Smith, M. A., \& Henry, G. 2002, ApJ, 575, 435R (RSH)

\bibitem[\protect\citeauthoryear{Saio}{2013}]{Saio2013}
Saio, H. 2013, Lect. Notes Phys. ``Seismology for Studies of stellar rotation,"
ed. K. Belkacem, 835, 3M.

\bibitem[\protect\citeauthoryear{Saio et al.}{2017}]{Saio et al.2017}
Saio, H., Ekstr\"om, S., Mowlavi, N. et al. 2017, MNRAS, 467, 3864S 

\bibitem[\protect\citeauthoryear{Saio}{2018}]{Saio2018}
Saio, H. 2018, Proc. Physics of Oscillating Stars Conf., arXiv:1812.01253

\bibitem[\protect\citeauthoryear{Semaan et al.}{2013}]{Semaan et al.2013}
   Semaan, T., Hubert, A. M., Zorec, J. et al 2013, A\&A, 551, 130SS 

\bibitem[\protect\citeauthoryear{Semaan et al.}{2018}]{Semaan et al.2018}
   Semaan, T., Hubert, A. M., Zorec, J. et al 2018, A\&A, 617, A70SS (Sem18)

\bibitem[\protect\citeauthoryear{Smith}{1995}]{Smith1995}
Smith, M. A. 1995, ApJ, 442, 812s

\bibitem[\protect\citeauthoryear{Smith}{2019a}]{Smith2019}
Smith, M. A. 2019, PASP, 131, 4201S (S19)

\bibitem[\protect\citeauthoryear{Smith et al.}{2006}]{Smith et al.2006}
Smith, M. A., Henry, G. W., \& Vishniac, E. 2006, ApJ, 647, 1375 (Paper\,1)

\bibitem[\protect\citeauthoryear{Smith et al.}{2012}]{Smith et al.2012}
Smith, M. A., Lopes de Oliveira, R., \& Motch, C., et al. 2012, A\&A, 540, 
A53S  (SLM)

\bibitem[\protect\citeauthoryear{Smith et al.}{2016}]{Smith et al.2016}
Smith, M. A., Oliveira, R., Motch, C.\ 2016, ASpR, 58, 782S (SLM2016) 

\bibitem[\protect\citeauthoryear{Smith \& Robinson}{1999}]{Smith&Robinson1999}
Smith, M. A., \& Robinson, R. D. 1999, ApJ, 517, 866S (SR99) 

\bibitem[\protect\citeauthoryear{Smith \& Robinson}{2003}]{Smith&Robinson2003}
Smith, M. A., \& Robinson, R. D. 2003, Interplay of periodic, cyclic, \& 
stochastic variability, ed. C. Sterken, ASP Conf. Ser. 292, 263S

\bibitem[Smith et al.(1998a)]{1998aApJ..503...877S}
Smith, M. A., Robinson, R. D., Corbet, R. H.\ 1998a, ApJ, 503, 877S (SRC)

\bibitem[\protect\citeauthoryear{Smith et al.}{1998b}]{Smith et al.1998b}
Smith, M. A., Robinson, R. D., Hatzes, A. P. 1998b, ApJ, 508, 945S (SRH)




\bibitem[\protect\citeauthoryear{Van\'{\i}\v cek}{2001}]{Vanicek1971}
Van\'{\i}\v cek, P. 1971, Ap\&SS, 12, 10V

\bibitem[\protect\citeauthoryear{Walker et al.}{2005}]{Walker et al.2005}
Walker, G. A., Kuschnig, R., Matthews, J. M., et al. 2005, ApJ, 635. L77W

\bibitem[\protect\citeauthoryear{Wang et al.}{2017}]{Wang et al.2017}
Wang L., Gies, D. R., \& Peters, G. J. 2017, ApJ, 843, 60W

\bibitem[\protect\citeauthoryear{Wang et al.}{2021}]{Wang et al.2021}
Wang L., Gies, D. R., \& Peters, G. J. et al. 2021, arXiv:2103.13642v1

\bibitem[\protect\citeauthoryear{White et al.}{1982}]{White et al.1982}
White, N. E., Swank, J. H., Holt, S. S. 1982, ApJ, 263, 277W

\bibitem[\protect\citeauthoryear{Yang et al.}{1988}]{Yang et al.1988}
Yang, S., Ninkov, Z., \& Walker, G. A. H. 1988, PASP, 100, 233Y


\end{thebibliography}
\end{document}